\documentclass[twoside,11pt]{article}

\usepackage{jsat_nologo}
\usepackage{amsmath, amsthm, amssymb}
\usepackage{pstricks}
\usepackage{pst-tree}
\usepackage{tikz}
\usepackage{enumitem}
\usepackage[ruled,vlined]{algorithm2e}

\input epsf

\ShortHeadings{A comparison of cardinality constraint encodings}
{E.J.W.Wynn}

\begin{document}
\tikzset{fontscale/.style = {font=\relsize{#1}} }
\newcommand{\Sat}{{\sc Sat} }

\title{A comparison of encodings for cardinality constraints in a SAT solver}

\author{\name Ed Wynn
	 \email ed.wynn@zoho.com \\
}
       
\maketitle

\begin{abstract}
Cardinality constraints are important in many \Sat problems; previous studies provide contradictory conclusions about the best encoding to use.  Here, three encodings are compared: Sinz's sequential-counter, Bailleux and Boufkhad's tree-based, and Ab{\'{\i}}o and coworkers' sort-based approaches.  The sequential-counter approach is found to be the fastest of these for a range of related, combinatorial test cases.  All encodings permit multiple solutions in the auxiliary variables for a single solution to the main variables; the numbers of multiple solutions can be very large, and might impede a \Sat solver.  Variants of the encodings are developed, where extra clauses reduce the numbers of multiple solutions.  These variants are found to have remarkably little effect on solution time, even when the number of clauses is approximately doubled.  The results accentuate the well-known observation that clause count and other measures of encoding size are not reliable indicators of the difficulty of a \Sat problem.
\end{abstract}


{\vspace{10pt}
 \relax{\noindent\footnotesize\baselineskip=9pt
 \hspace{20pt}{\em 27 October 2018.}}}

\section{Introduction}
\label{sec:intro}

We will use the notation $\leqslant_r\!\left( x_1, \dots, x_n \right)$ for the \emph{cardinality constraint}:
\begin{equation}\label{cardinal1}
\displaystyle\sum_{i=1}^{n}{x_i} \leqslant r
\end{equation}
Such constraints are frequently required in \Sat problems.  In some \Sat solvers, these constraints may be handled natively, and this may be efficient~\cite{Manthey:2013}\cite{Maglalang:2012}.  In general, however, it is required to encode a constraint into standard clauses in Conjunctive Normal Form. Several encodings have been proposed, as discussed below.  Typical approaches introduce \emph{auxiliary variables}, in addition to the \emph{main variables}, $\{x_i\}$.

Previous studies have compared results from various encodings.  The diversity of their conclusions is one motivation for the current study:
\begin{itemize}
\item Frisch and Giannoros~\cite{Frisch:2010} obtained good results from Sinz's sequential-counter encoding; for example, in comparison with the Commander encoding, which they extended to fit the test problem.  Bailleux and Boufkhad's tree-based encoding was rejected from consideration.
\item Marques-Silva and Lynce~\cite{MarquesSilva:2007} used Sinz's sequential-counter encoding to apply many $\leqslant_1$ constraints per problem.  This was compared with the `naive', pairwise method for $\leqslant_1$ constraints.  The pairwise method is often discounted as impractical~\cite{Chen:2010}, because it generates a quadratic number of clauses, or is found in practice to be slow~\cite{Nguyen:2015}.  In~\cite{MarquesSilva:2007}, however, the sequential-counter method was slower than the pairwise method, and showed much more variability, unless solver modifications were made.
\item Knuth~\cite{Knuth:2015} presented versions of Sinz's sequential-counter and the Bailleux-Boufkhad tree-based encodings.  On the basis of extensive testing, the latter was preferred.
\item Martins~\cite{Martins:2013} compared Sinz's sequential-counter encoding, the Bailleux-Boufkhad tree-based encoding and an enhanced version of E\'en and S\"orensson's sorting-network encoding, in the context of MaxSAT test cases.  These were found to be in increasing order of speed.
\end{itemize}

The present study compares three encodings with substantially different approaches: the sequential counter of Sinz; the tree-based approach of Bailleux and Boufkhad; and the sorting-network approach of Ab{\'{\i}}o and coworkers.  (A related method due to Jabbour and coworkers is shown to be equivalent to the Sinz sequential-counter encoding.)  However, before this comparison, the encodings are reviewed, in particular to develop variants to the encodings.  We observe that the encodings, as presented, typically allow multiple solutions in the auxiliary variables for a single solution in the main variables.  These unnecessary solutions might impede a \Sat solution.  Therefore, ways to reduce or remove the multiple solutions are developed here; this is here called \emph{strengthening}.  The runtime comparison then investigates whether the strengthened problem can be solved more quickly, for example because the unnecessary solutions have been removed, or alternatively whether the strengthening clauses do not justify their extra expense.

\section{Encodings for cardinality constraints}
\label{sec:encodings}

\subsection{Sequential-counter encoding}
\label{sec:sinz}
Sinz~\cite{Sinz:2005} presents an encoding of $\leqslant_r\!\left( x_1, \dots, x_n \right)$.  This is based on a sequential counting circuit.  (In the same paper, Sinz presents an alternative encoding based on a parallel counter.  This is characterised by a low clause count, but it has been found to be slow~\cite{BenHaim:2012} and it is not considered further here.)  In this section and the next, we will consider alternative ways to understand this encoding, and use these in Section~\ref{sec:sinzstrengthening} to strengthen the encoding.  The original presentation was open to immediate simplification (containing, for example, some clauses with only single literals).  Therefore, the exposition of Knuth~\cite{Knuth:2015} is used here, with some variations.

We will consider a staggered grid of auxiliary variables $e_j^k$ with $1 \leqslant k \leqslant r$ and $k \leqslant j \leqslant n+k-r-1$.  (This staggered grid is a difference from Knuth's exposition, in notation only. An example of the staggered grid is shown later, in Figure~\ref{fig:sinzexamples}.)  The clauses encoding the constraint are:
\begin{equation}\label{eq:sinz1}
\bigwedge_{k=1}^r \bigwedge_{j=k}^{n+k-r-2} {\lnot e_j^k \lor e_{j+1}^k}
\end{equation}
\begin{equation}\label{eq:sinz2}
\bigwedge_{k=0}^r \bigwedge_{j=k}^{n+k-r-1} {\lnot e_j^k \lor e_{j+1}^{k+1} \lor \lnot x_{j+1}}
\end{equation}
except that $e_j^0$ is implicitly true and $e_j^{r+1}$ is implicitly false for all $j$; these variables should be omitted from Eq.~\ref{eq:sinz2}.  The number of auxiliary variables is $r (n-r)$.

The following condition applies to the auxiliary variables when the clauses in Eq.~\ref{eq:sinz1} and Eq.~\ref{eq:sinz2} are applied:
\begin{equation}\label{eq:sinzcondition}
e_j^k=1 \ \text{whenever}\  \sum_{i=1}^{j} x_i \geqslant k
\end{equation}
Conversely, this condition can be used to derive the clauses (see Exercise~26 in~\cite{Knuth:2015}).  This explains why the staggered grid is appropriate: When $k>j$, the condition cannot be satisfied, so the variable is not needed.  For sufficiently large $j$ and small $k$, namely $j\geqslant n-r+k$, it does not matter whether the condition is true: if so many main variables are false, the constraint is automatically satisfied and no further action is required.

It is important to note that the condition, Eq.~\ref{eq:sinzcondition}, defines when $e_j^k$ must be true, but in general the auxiliary variables can also be true when the condition does not require it.  This is discussed in Section~\ref{sec:sinzstrengthening}.

The staggered grid can be squared up by moving rows to the left: with $s_j^k = e_{j+k-1}^k$, $1 \leqslant k \leqslant r$ and $1 \leqslant j \leqslant n-r$, the clauses can be restated:
\begin{equation}\label{eq:sinzK1}
\bigwedge_{k=1}^r \bigwedge_{j=1}^{n-r-1} {\lnot s_j^k \lor s_{j+1}^k}
\end{equation}
\begin{equation}\label{eq:sinzK2}
\bigwedge_{k=0}^r \bigwedge_{j=1}^{n-r} {\lnot s_j^k \lor s_{j}^{k+1} \lor \lnot x_{j+k}}
\end{equation}
similarly omitting  $s_j^0$ and $s_j^{r+1}$.  This is Knuth's statement of the clauses~\cite{Knuth:2015}.

\subsection{Enhanced Pigeonhole Encoding}
\label{sec:jabbour}
Jabbour and coworkers~\cite{Jabbour:2013} present an encoding for $\geqslant_q\!\left( x_1, \dots, x_n \right)$ based on pigeonhole principles.  This can be converted to apply to $\leqslant_r\!\left( x_1, \dots, x_n \right)$, which is identical in effect to $\geqslant_{n-r}\!\left( \lnot x_1, \dots, \lnot x_n \right)$.  Since this approaches cardinality from another direction, the question arises whether this encoding has different characteristics to the Sinz encoding, for example.  This section answers that question.

Jabbour and coworkers start by considering (and correcting) an encoding of $\geqslant_q\!\left( x_1, \dots, x_n \right)$ reported by Warners and attributed to Hooker.  The encoding uses a block of $q \times n$ auxiliary variables, $z_j^k$.  Clauses are developed to encode the following:
\begin{itemize}[noitemsep,topsep=0pt]
\item  If $z_j^k$ is true for any $k$, then the main variable $x_j$ must be true.
\item For any row $k$, at least one $z_j^k$ must be true.
\item For any column $j$, at most one $z_j^k$ can be true.
\end{itemize}

Jabbour and coworkers point out that, while this encoding clearly enforces the constraint as required, it allows unnecessary freedom in the solution.  For example, the rows of auxiliary variables in a solution can be permuted, and the result will still be a solution.  Starting from a solution with the minimum number $q$ of true main variables, further main variables may be set to true, either without changing the auxiliary variables or by making auxiliary variables true in any row of the appropriate columns.  Jabbour and coworkers note that these freedoms might be detrimental to speed of \Sat solving: for example, to demonstrate unsatisfiability, the solver must consider all the permutations.

Jabbour and coworkers remove some freedoms in the encoding by requiring that the rows of auxiliary variables should be ordered.  We here restate their clauses with a change of notation, which produces a considerable simplification but does not change the effective clauses.  The staggered grid of auxiliary variables is now $f_j^k$ with $1\leqslant k\leqslant q$ and $k \leqslant j\leqslant n+k-q$.  The clauses are:
\begin{equation}\label{eq:jabbour1}
\bigwedge_{k=1}^q \bigwedge_{j=k}^{k+n-q} {\lnot f_j^k \lor x_j}
\end{equation}
\begin{equation}\label{eq:jabbour2}
\bigwedge_{k=1}^q \bigvee_{j=k}^{k+n-q} f_j^k
\end{equation}
\begin{equation}\label{eq:jabbour3}
\bigwedge_{k=1}^{q-1} \bigwedge_{j=k+1}^{k+n-q} {\left( \lnot f_j^k \lor \bigvee_{i=j+1}^{k+n-q+1} f_j^{k+1} \right)}
\end{equation}
Eq.~\ref{eq:jabbour3} removes the `at most one' clauses in the original encoding, and instead requires that if $f_j^k$ is true then one of the $f_i^{k+1}$ must also be true, with $i>j$.  Therefore, the rightmost entries in the rows are in strictly increasing order.  A small extra freedom has arrived: if $f_j^k$ is true, then $f_j^{k+a}$ with positive $a$ may be true or false without affecting any other variables.

This encoding improves on the original encoding, which used the naive binomial encoding of `at most one' constraints.  The number of auxiliary variables is $\Theta(r (n-r))$, as for Sinz's encoding.  However, the clauses in Eq.~\ref{eq:jabbour3} are relatively long, and the number of literals in all clauses is $\Theta(r^2 (n-r))$.  So, here we develop a different encoding on the same principles.

Instead of auxiliary variables $f_j^k$ that imply whether $x_j$ must be true, we here propose auxiliary variables $g_j^k$ such that a \emph{transition} from $g_j^k = 1$ to $g_{j+1}^k = 0$ implies that $x_{j+1}$ must be true.  We implicitly set $g_{k-1}^k = 1$ and $g_{n+k-q}^k = 0$, so that there is inevitably at least one transition in each row.  This removes the need for clauses equivalent to Eq.~\ref{eq:jabbour2}.  It also saves one auxiliary variable per row: we need the staggered grid of $g_j^k$ for $1\leqslant k\leqslant q$, $k\leqslant j\leqslant n+k-q-1$.  The clauses are then:
\begin{equation}\label{eq:jabbourE1}
\bigwedge_{k=1}^{q-1} \bigwedge_{j=k}^{n+k-q-1} {\lnot g_j^k \lor g_{j+1}^{k+1}}
\end{equation}
\begin{equation}\label{eq:jabbourE2}
\bigwedge_{k=1}^q \bigwedge_{j=k-1}^{n+k-q-1} {\lnot g_j^k \lor g_{j+1}^{k} \lor x_{j+1}}
\end{equation}
Eq.~\ref{eq:jabbourE1} orders the transitions: any true $g_j^k$ implies true $g_{j+1}^{k+1}$, so the rightmost transitions in the rows are in strictly increasing order. Eq.~\ref{eq:jabbourE2} requires that a transition implies that the corresponding main variable is true.  In these clauses, we omit $g_{k-1}^k$ and $g_{n+k-q}^{k+1}$.

As we did for the Sinz encoding, we can square up the staggered grid, defining $t_j^k = g_{j+k-1}^k$ with $1\leqslant k\leqslant q$, $1\leqslant j\leqslant n-q$.  The rearranged clauses are:
\begin{equation}\label{eq:jabbourEE1}
\bigwedge_{k=1}^{q-1} \bigwedge_{j=1}^{n-q} {\lnot t_j^k \lor t_{j}^{k+1}}
\end{equation}
\begin{equation}\label{eq:jabbourEE2}
\bigwedge_{k=1}^q \bigwedge_{j=0}^{n-q} {\lnot t_j^k \lor t_{j+1}^{k} \lor x_{j+k}}
\end{equation}
We can now substitute $q=n-r$ and replace main variables with their complements, so that the clauses enforce the same $\leqslant_r$ constraint as in the previous section. After these changes, and when the squared-up grid is transposed ($s_j^k = t_k^j$) and $j$ and $k$ are swapped, we observe something remarkable: \emph{the clauses in Eq.~\ref{eq:jabbourEE1} and Eq.~\ref{eq:jabbourEE2} are identical to the clauses in Eq.~\ref{eq:sinzK1} and Eq.~\ref{eq:sinzK2}}.  So, this gives a completely negative answer to the question of whether the approach in this section has different characteristics to the sequential-counter encoding: the two approaches produce identical clauses (when the pigeonhole approach is encoded with transition-based auxiliary variables).

\subsection{Strengthening the sequential-counter encoding}
\label{sec:sinzstrengthening}

The previous section demonstrated that there is no need to consider the pigeonhole encoding (using transition-based auxiliary variables) separately from Sinz's sequential-counter encoding.  However, the two derivations of the clauses give different insights, allowing the method to be visualised and adapted.  In particular, Section~\ref{sec:jabbour} considers local patterns and transitions in the auxiliary variables, whereas Section~\ref{sec:sinz} is based on Eq.~\ref{eq:sinzcondition}, which relates each auxiliary variable to a sum of main variables.  In this section, we consider the possibility of multiple solutions in the auxiliary variables for a single ordered set of values of the main variables.  We then develop clauses that can reduce the number of these solutions.  

Suppose that the values of the main variables have been fixed, such that the $\leqslant_r$ constraint is satisfied.  We define the \emph{canonical solution} to the auxiliary variables in the staggered grid of Section~\ref{sec:sinz}:
\begin{itemize}[noitemsep,topsep=0pt]
\item  If $x_j$ is the $k$-th true main variable, then the $k$-th row of auxiliary variables has a transition from $e_{j-1}^k = 0$ to $e_{j}^k = 1$.
\item Eq.~\ref{eq:sinz1} requires that at most one transition occurs in each row: $e_i^k = 0$ for all $i<j$, and $e_i^k = 1$ for all $i\geqslant j$.  If the first auxiliary variable in a row, $e_k^k$, is 1, then this is implicitly a transition from $e_{k-1}^k=0$.
\item If the number of true main variables is $i$, then all auxiliary variables $e_j^k=0$ for $k>i$.
\end{itemize}

As discussed below, the canonical solution is compatible with Eq.~\ref{eq:sinz1} and Eq.~\ref{eq:sinz2}, and a strengthened version of Eq.~\ref{eq:sinzcondition}:
\begin{equation}\label{eq:sinzconditionstrong}
e_j^k=1 \ \text{iff}\  \sum_{i=1}^{j} x_i \geqslant k
\end{equation}

\begin{figure}[tp] 
\centering
\begin{tikzpicture}
\definecolor{mygrey}{rgb}{0.7,0.7,0.7}
\definecolor{mymidblue}{rgb}{0.1718,0.4531,0.7031}
\definecolor{mydarkblue}{rgb}{0.0859,0.2265,0.3515}

\draw (-1,-1/2) node{$k=$};
\foreach \y in {1,...,4} { \draw (-1/2,-\y/2) node{$\y$}; }
\draw (0,1/2) node{$j=$};
\foreach \x in {1, ..., 10} { \draw(\x/2,1/2) node{$\x$}; }

\foreach \x in {0, ..., 5} { \draw(\x/2,0/2) node{\footnotesize \textcolor{mymidblue}{(1)}};     \draw [thin][mymidblue] (\x/2,0/2) +(-.25,-.21) rectangle ++(.25,.29); }
\foreach \pos in {(-0/2,-1/2),(1/2,-2/2),(2/2,-3/2),(3/2,-4/2)} { \draw \pos node{\footnotesize \textcolor{mymidblue}{(0)}};  \draw [thin][mymidblue] \pos +(-.25,-.21) rectangle ++(.25,.29);}
\foreach \x in {5, ..., 10} { \draw(\x/2,-5/2) node{\footnotesize \textcolor{mymidblue}{(0)}};  \draw [thin][mymidblue] (\x/2,-5/2) +(-.25,-.21) rectangle ++(.25,.29); }

\foreach \y [evaluate=\y as \shiftedy using {int(0+\y)}] [evaluate=\y as \xmin using {int(\y)}] [evaluate=\y as \xmax using {int(\y+5)}] in {1,...,4} {
  \foreach \x [evaluate=\x as \shiftedx using {int(0+\x)}] in {\xmin, ...,\xmax} {
    \draw [ultra thick](\shiftedx/2,-\shiftedy/2) +(-.25,-.21) rectangle ++(.25,.29);
}}

\draw (14/2,1/2) node{$j=$};
\foreach \x [evaluate=\x as \shiftedx using {int(14+\x)}] in {1, ..., 10} { \draw(\shiftedx/2,1/2) node{$\x$}; }

\foreach \y [evaluate=\y as \shiftedy using {int(0+\y)}] [evaluate=\y as \xmin using {int(\y)}] [evaluate=\y as \xmax using {int(\y+5)}] in {1,...,4} {
  \foreach \x [evaluate=\x as \shiftedx using {int(14+\x)}] in {\xmin, ...,\xmax} {
    \draw [ultra thick](\shiftedx/2,-\shiftedy/2) +(-.25,-.21) rectangle ++(.25,.29);
}}
\foreach \y [evaluate=\y as \shiftedy using {int(6+\y)}] [evaluate=\y as \xmin using {int(\y)}] [evaluate=\y as \xmax using {int(\y+5)}] in {1,...,4} {
  \foreach \x [evaluate=\x as \shiftedx using {int(0+\x)}] in {\xmin, ...,\xmax} {
    \draw [ultra thick](\shiftedx/2,-\shiftedy/2) +(-.25,-.21) rectangle ++(.25,.29);
}}
\foreach \y [evaluate=\y as \shiftedy using {int(6+\y)}] [evaluate=\y as \xmin using {int(\y)}] [evaluate=\y as \xmax using {int(\y+5)}] in {1,...,4} {
  \foreach \x [evaluate=\x as \shiftedx using {int(14+\x)}] in {\xmin, ...,\xmax} {
    \draw [ultra thick](\shiftedx/2,-\shiftedy/2) +(-.25,-.21) rectangle ++(.25,.29);
}}

\foreach \xstart [count=\y] [evaluate=\y as \xmax using {int(\y+5)}]  [evaluate=\y as \shiftedy using {int(0+\y)}] in {3,6}  {
 \foreach \x [evaluate=\x as \shiftedx using {int(0+\x)}] in {\xstart, ...,\xmax} {
   \draw [fill=mygrey,ultra thick](\shiftedx/2,-\shiftedy/2) +(-.25,-.21) rectangle ++(.25,.29);
   \draw(\shiftedx/2,-\shiftedy/2) node{1};
}}
\foreach \xstart [count=\y] [evaluate=\y as \xmax using {int(\y+5)}]  [evaluate=\y as \shiftedy using {int(0+\y)}] in {1,3,5,6}  {
 \foreach \x [evaluate=\x as \shiftedx using {int(14+\x)}] in {\xstart, ...,\xmax} {
   \draw [fill=mygrey,ultra thick](\shiftedx/2,-\shiftedy/2) +(-.25,-.21) rectangle ++(.25,.29);
   \draw(\shiftedx/2,-\shiftedy/2) node{1};
}}
\foreach \xstart [count=\y] [evaluate=\y as \xmax using {int(\y+5)}]  [evaluate=\y as \shiftedy using {int(6+\y)}] in {3,3,6,6}  {
 \foreach \x [evaluate=\x as \shiftedx using {int(0+\x)}] in {\xstart, ...,\xmax} {
   \draw [fill=mygrey,ultra thick](\shiftedx/2,-\shiftedy/2) +(-.25,-.21) rectangle ++(.25,.29);
   \draw(\shiftedx/2,-\shiftedy/2) node{1};
}}
\foreach \xstart [count=\y] [evaluate=\y as \xmax using {int(\y+5)}]  [evaluate=\y as \shiftedy using {int(6+\y)}] in {3,6,4,6}  {
 \foreach \x [evaluate=\x as \shiftedx using {int(14+\x)}] in {\xstart, ...,\xmax} {
   \draw [fill=mygrey,ultra thick](\shiftedx/2,-\shiftedy/2) +(-.25,-.21) rectangle ++(.25,.29);
   \draw(\shiftedx/2,-\shiftedy/2) node{1};
}}

\end{tikzpicture}
\caption{Examples of solutions for the auxiliary variables $e_j^k$ to Eq.~\ref{eq:sinz1} and Eq.~\ref{eq:sinz2}, the unstrengthened sequential-counter encoding of  $\leqslant_r\!\left( x_1, \dots, x_n \right)$, with $n=10$, $r=4$, $x_3=x_6=1$ and other $x_i=0$.  Blue text in parentheses shows implicit values.  Empty boxes imply the auxiliary variable is 0.  Top left: the canonical solution. Top right: a solution compatible with Eq.~\ref{eq:sinz3} but not Eq.~\ref{eq:sinz4}; this is the canonical solution for $x_1=x_3=x_5=x_6=1$.  Bottom left: a solution compatible with Eq.~\ref{eq:sinz4} but not Eq.~\ref{eq:sinz3}. Bottom right: a solution that would be rejected by Eq.~\ref{eq:sinz3} and/or Eq.~\ref{eq:sinz4}.}
\label{fig:sinzexamples}
\end{figure}
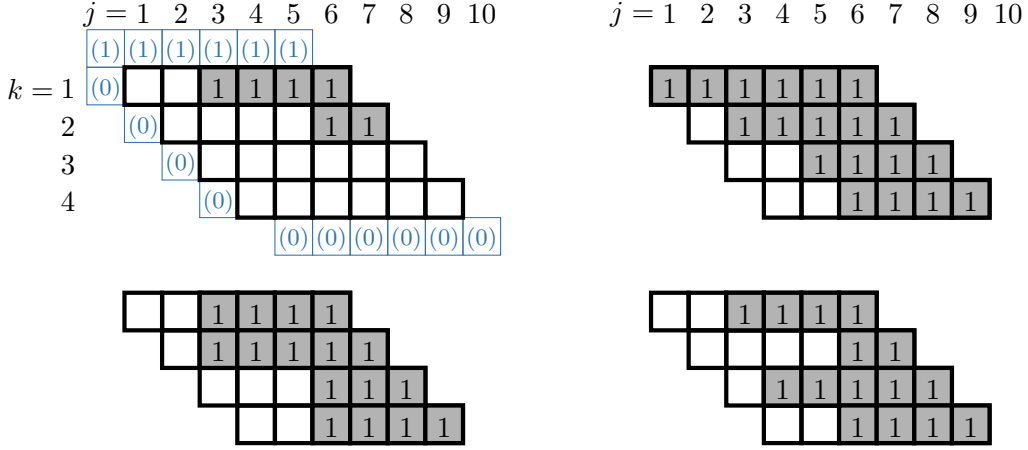

When there are $r$ true main variables, so that the constraint is tight, then the canonical solution is the only one.  However, if there are fewer true main variables, many other solutions are possible.  For example, if $x_j=1$, then transitions from $e_{j-1}^k = 0$ to $e_{j}^k = 1$ can occur for several values of $k$.  Transitions can also occur for any other value of $j$, except that the next true main variable may then cause more than one transition.  Some examples are shown in Figure~\ref{fig:sinzexamples}. The clearest and most extreme case is when all the main variables are false; in this case, each row can contain up to one transition at any position.  There are $(n-r)^r$ solutions of the auxiliary variables for this case.

Further clauses can be imposed, which remove some of the solutions but are still compatible with the canonical solution:
\begin{equation}\label{eq:sinz3}
\bigwedge_{k=1}^{r-1} \bigwedge_{j=k}^{n+k-r-1} {e_j^k \lor \lnot e_{j+1}^{k+1}}
\end{equation}
\begin{equation}\label{eq:sinz4}
\bigwedge_{k=1}^r \bigwedge_{j=k-1}^{n+k-r-2} {e_j^k \lor \lnot e_{j+1}^k \lor x_{j+1}}
\end{equation}
In Eq.~\ref{eq:sinz4}, $e_{k-1}^k$ is implicitly false and is omitted.  These new clauses can be understood geometrically or by reference to the strengthened condition Eq.~\ref{eq:sinzconditionstrong}:
\begin{itemize}
\item Geometrically, Eq.~\ref{eq:sinz3} requires that blank squares propagate to the south-east ($e_j^k=0$ forces $e_{j+1}^{k+1}=0$) and filled squares to the north-west ($e_{j+1}^{k+1}=1$ forces $e_{j}^k=1$).
\item In Eq.~\ref{eq:sinzconditionstrong}, if the sum up to $x_j$ is less than $k$, then the sum up to $x_{j+1}$ must be less than $(k+1)$, and Eq.~\ref{eq:sinz3} follows.
\item From both explanations of Eq.~\ref{eq:sinz3}, we conclude that when these extra clauses are imposed, every solution to the auxiliary variables is the canonical solution to some ordered set of main-variable values: the essential property of a canonical solution is that $0\rightarrow 1$ transitions occur in strictly increasing positions, row by row.  However, other main-value solutions with fewer true values are also compatible with each canonical solution.  (See the example top-right in Figure~\ref{fig:sinzexamples}.)
\item Geometrically, Eq.~\ref{eq:sinz4} requires that a $0\rightarrow 1$ transition in a row is necessarily associated with a true main variable.  If $x_j=0$, then false auxiliary values propagate to the east ($e_j^k=0$ forces $e_{j+1}^k=0$) and true values propagate to the west ($e_{j+1}^k=1$ forces $e_{j}^k=1$).
\item In terms of Eq.~\ref{eq:sinzconditionstrong}, Eq.~\ref{eq:sinz4} represents two facts: the sum of main variables can only increase with a true main variable; and if the main variable is false, then the sum does not change.
\end{itemize}

Eq.~\ref{eq:sinz1} and Eq.~\ref{eq:sinz2} will be called the \emph{unstrengthened} sequential-counter encoding; when Eq.~\ref{eq:sinz3} and Eq.~\ref{eq:sinz4} are added, the \emph{fully strengthened} encoding.  When the fully strengthened encoding is imposed, only the canonical solution is compatible with an ordered set of main-variable values.  This can be shown by induction on the rows: Assume that row $(k-1)$ has a $0\rightarrow 1$ transition in the canonical location.  Row $k$ will not have a transition at an earlier or equal position, because of Eq.~\ref{eq:sinz3}.  Nor will it have a transition before the next true main variable, because of Eq.~\ref{eq:sinz4} -- and this also applies to $k=1$, because all $e_j^0$ are implicitly true.  Row $k$ will have a transition at the next true main variable, because of Eq.~\ref{eq:sinz2}, and the true values then fill the rest of the row by Eq.~\ref{eq:sinz1}.

The fully strengthened sequential-counter encoding for the constraint $\leqslant_r\!\left( x_1, \dots, x_n \right)$ can be converted to the equality constraint, $=_r\!\left( x_1, \dots, x_n \right)$, by changing the upper limit of $j$ in Eq.~\ref{eq:sinz4} to $(n+k-r-1)$, with $e_{n+k-r}^k$ implicitly true for all $k$.  Essentially, these end values change from being irrelevant to representing the last possible position for the $k$-th true main variable.

Compared to the unstrengthened encoding, the fully strengthened encoding has no extra auxiliary variables, but approximately twice as many clauses.  The number of extra clauses is $2r(n-r)-n+r$, which is sometimes less than and sometimes more than the number of unstrengthened clauses, $2r(n-r)+n-2r$.

\subsection{Tree-based encoding}
\label{sec:bail}

Bailleux and Boufkhad~\cite{Bailleux:2003} present an encoding of $\leqslant_r\!\left( x_1, \dots, x_n \right)$ based on a tree of variable counts.  As for the Sinz paper, more than one formulation is presented, and other workers have noted improvements --  see for example \cite{Buttner:2005}.  Without these improvements, the approach has been rejected in other studies, for example~\cite{Frisch:2010}.  Knuth~\cite{Knuth:2015} presents a coherent version, which is used here.  However it is noted (in Exercise~24 of~\cite{Knuth:2015}) that this version still contains some inefficient aspects: some pure literals are introduced.  (A pure literal is present in the clauses in only one polarity -- its negation is never required.  Therefore, that literal can be made true, and all clauses involving it will be automatically satisfied.  Solvers can detect and remove pure literals during preprocessing of clauses, so the inefficiency is presumably small, but it is avoidable.)  The encodings generated by Knuth's code (available in the {\sc{SATexamples}} package~\cite{Knuth:web}) sometimes have other inefficiencies, such as unary clauses in encodings of $=_r$ constraints.  Therefore, an improved presentation is given as pseudocode in Figure~\ref{fig:bail}, but the overall structure is taken directly from~\cite{Knuth:2015}.

The algorithm places the main variables as leaves in a binary tree.  This can be packed into an array of size $2n-1$, where the leaves have indices $n$ to $2n-1$ and any other node $k$, for $1\leqslant k < n$, has daughters $2k$ and $2k+1$.  For each node $k$, there are auxiliary variables $b_j^k$ for some values of $j$.  The clauses produced by Figure~\ref{fig:bail} ensure that $b_j^k$ obeys the condition:
\begin{equation}\label{eq:bailcondition}
b_j^k=1 \ \text{whenever the leaves below node}\  k\ \text{contain}\ j\ \text{or more 1s}
\end{equation}
This is  similar to the condition in  Eq.~\ref{eq:sinzcondition} for the sequential-counter encoding, in the following sense: the auxiliary variables are permitted to adopt true values even when the condition does not require them.  Therefore, this encoding can be strengthened, as discussed in Section~\ref{sec:bailstrengthening}.

The only remaining issue is to write clauses to restrict the counts $b_j^k$ to enforce the $\leqslant_r$ constraint (in the first nested loops of Figure~\ref{fig:bail}), and to define each count in terms of its daughters' counts (the second nested loops).  The number of leaves below node $k$ is $L_k$, and the count for each subtree must obey $\leqslant_r$, so $t_k=\min(r,L_k)$ is the maximum count that needs to be considered at node $k$.  Figure~\ref{fig:bail} represents a small improvement in noting only the variables that are required.  Because the nodes are visited in top-down order in the second nested loops, the requirements cascade down, and only one pass is needed.

	\begin{figure}[ht] 
	\centering
\setlength{\interspacetitleruled}{0pt} 
\setlength{\algotitleheightrule}{0pt} 
\begin{algorithm}[H]
\SetAlgoNoLine
\SetAlgoNoEnd
\DontPrintSemicolon
\KwResult{Clauses are written to enforce  $\leqslant_r\!\left( x_1, \dots, x_n \right)$  }
  Note: The variable $b_0^k$ for any $k$ is implicitly true, so $\lnot b_0^k$ can be omitted from any clause where it appears, and the variable is not flagged as required.\;
  Note: The variable $b_1^k$ for $n\leqslant k <2n$ is replaced by the main variable $x_{k-n+1}$.\;
  \For{$k=2 n-1$ \KwTo $n$} {
    $L_k$=1, $t_k$=1
  }
  \For{$k=n-1$ \KwTo $1$}{
    $L_k = L_{2k} + L_{2k+1}$, $t_k = \min(r,L_k)$
  }
  \For{$k=1$ \KwTo $n-1$} {
    \For{$i=1$ \KwTo $t_{2k}$ } {
      \For{$j=1$ \KwTo $t_{2k+1}$} {
          \uIf{$i+j == r+1$} {
             Write clause $\left(\lnot b_i^{2k} \lor \lnot b_j^{2k+1}\right)$\;
             and note that $b_i^{2k}$ and $b_j^{2k+1}$ are required.
        }
      }
    }
  }
  \For{$k=2$ \KwTo $n-1$} {
    \For{$m=1$ \KwTo $t_k$} {
      \uIf{ $b_m^k$ is required } {
        \For{$i=0$ \KwTo $t_{2k}$} {
          \For{$j=0$ \KwTo $t_{2k+1}$} {
            \uIf{$i+j == m$} {
              Write clause $\left(\lnot b_i^{2k} \lor \lnot b_j^{2k+1} \lor b_m^k \right)$\;
              and note that $b_i^{2k}$ and $b_j^{2k+1}$ are required.
              }
            }
          }
        }
      }
    }
\end{algorithm}

	\caption{Pseudocode to generate the Bailleux-Boufkhad tree-based encoding of $\leqslant_r$.}
	\label{fig:bail}
	\end{figure} 

\subsection{Strengthening the tree-based encoding} \label{sec:bailstrengthening}

As Knuth shows (in Exercise~30~\cite{Knuth:2015}), the $\leqslant_r$ constraint in Figure~\ref{fig:bail} can be converted to an $=_r$ constraint without extra auxiliary variables.  Essentially, exactly the same procedure is applied to create an encoding of $\geqslant_{n-r}\!\left( \lnot x_1, \dots, \lnot x_n \right)$, but the auxiliary variables are reused.  Pseudocode is presented in Figure~\ref{fig:bailexact}.  When the resulting additional clauses are applied, the auxiliary variables obey a stronger condition:
\begin{equation}\label{eq:bailconditionstrong}
b_j^k=1 \ \text{iff the leaves below node}\  k\ \text{contain}\ j\ \text{or more 1s}
\end{equation}
If the auxiliary variables obey this condition, they will be said to have \emph{canonical values}.

The $=_r$ constraint is more than a strengthened version of the $\leqslant_r$ constraint, because it restricts the main-variable solutions.  However, the procedure of Figure~\ref{fig:bailexact} can be used to strengthen the $\leqslant_r$ constraint, simply by omitting any clause that restricts a main variable.  This still encodes only the $\leqslant_r$ constraint, because there are no clauses at all involving main variables in positive polarity.  Only the auxiliary variables have been constrained, and the canonical condition of Eq.~\ref{eq:bailconditionstrong} is compatible with the $\leqslant_r$ constraint.  Using the reduced version of Figure~\ref{fig:bailexact} will be called \emph{inequality strengthening} of the Bailleux-Boufkhad encoding.

Another form of strengthening will be called \emph{sideways strengthening}: at any non-leaf node $k$, clauses $\left(b_i^k \lor \lnot b_{i+1}^k \right)$ can be generated for any $i$ where both those variables are already required.  Each such clause expresses the fact that whenever the leaves below node $k$ contain $i+1$ or more 1s, then they must also contain $i$ or more 1s.  This fact will be present in the solution once all the variables have been assigned values consistent with each other and Eq.~\ref{eq:bailcondition}.  However, it is not a fact that is immediately enforced by the unstrengthened clauses in the encoding (so-called \emph{unit propagation}).  Therefore, it is possible that sideways-strengthening will assist the generation of a solution.  Each clause also has a converse effect: if $b_{i}^k=0$, which is possible only if the leaves below $k$ do not contain as many as $i$ 1s, then $b_{i+1}^k$ is also forced to be 0.  This is not imposed by the unstrengthened clauses; the unstrengthened condition, Eq.~\ref{eq:bailcondition}, places no restrictions on when auxiliary variables must be false.  Therefore, this sideways strengthening can reduce the number of solutions of auxiliary variables that are compatible with an ordered set of main-variable values.  Sideways strengthening and inequality strengthening can be applied together or independently; the costs and benefits of this are explored in Section~\ref{sec:compare}.

In Section~\ref{sec:sinzstrengthening}, it was noted that, for the unstrengthened sequential-counter encoding, there are $(n-r)^r$ ordered sets of auxiliary-variable values that are possible when all the main variables are false.  Essentially, when an auxiliary variable adopts a true value that is not required by the canonical solution, it consititutes a `false alarm', which unnecessarily constrains other variables (auxiliary and/or main).  The all-zero set of main variables is the one that is most liable to allow false alarms and multiple solutions.  The same position applies for the unstrengthened Bailleux-Boufkhad encoding.  The number of solutions is more difficult to quantify, but it can again be very substantial.  For example, if we take any ordered set of values of the main values and calculate the canonical values of the auxiliary variables, then those auxiliary variables will still be compatible with the unstrengthened clauses if we change some true main variables to false.  Hence, a lower bound on the number of solutions compatible with the all-zero mains is the sum of the binomials: $\binom{n}{0} + \cdots + \binom{n}{r}$.  However, in all these solutions, the false alarms start at the nodes adjacent to leaves; in general, many other solutions are possible.

	\begin{figure}[ht] 
	\centering
\setlength{\interspacetitleruled}{0pt} 
\setlength{\algotitleheightrule}{0pt} 
\begin{algorithm}[H]
\SetAlgoNoLine
\SetAlgoNoEnd
\DontPrintSemicolon
\KwResult{Building on Figure~\ref{fig:bail}, extra clauses are written to enforce  $=_r\!\left( x_1, \dots, x_n \right)$  }
  \For{$k=1$ \KwTo $2n-1$}{
    $u_k = \min(n-r,L_k)$
  }
  \For{$k=1$ \KwTo $n-1$} {
    \For{$i=1$ \KwTo $u_{2k}$ } {
      \For{$j=1$ \KwTo $u_{2k+1}$} {
          \uIf{$i+j == n-r+1$} {
             Write clause $\left( b_{L_{2k}+1-i}^{2k} \lor b_{L_{2k+1}+1-j}^{2k+1}\right)$\;
             and note that these variables are required in the extra clauses.
        }
      }
    }
  }
  \For{$k=2$ \KwTo $n-1$} {
    \For{$m=1$ \KwTo $u_k$} {
      \uIf{ $b_{L_k+1-m}^k$ is required in the extra clauses} {
        \For{$i=0$ \KwTo $u_{2k}$} {
          \For{$j=0$ \KwTo $u_{2k+1}$} {
            \uIf{$i+j == m$} {
              Write clause $\left( b_{L_{2k}+1-i}^{2k} \lor b_{L_{2k+1}+1-j}^{2k+1} \lor \lnot b_{L_k+1-m}^k\right)$\;
              and note that these variables are required in the extra clauses.
              }
            }
          }
        }
      }
    }
\end{algorithm}

	\caption{Additional pseudocode to convert the encoding in Figure~\ref{fig:bail} to $=_r$.}
      \label{fig:bailexact}
	\end{figure} 

\subsection{Sort-based encoding}
\label{sec:abio}

\begin{figure}[ht] 
	\centering

\begin{tikzpicture}
\definecolor{mygrey}{rgb}{0.7,0.7,0.7}
\definecolor{mymidblue}{rgb}{0.1718,0.4531,0.7031}
\definecolor{mydarkblue}{rgb}{0.0859,0.2265,0.3515}
\definecolor{mymidred}{rgb}{0.8359,0.1875,0.1484}
\definecolor{mydarkred}{rgb}{0.4179,0.0938,0.0742}

\foreach \i in {1,...,9} {
  \draw [very thick](2.1/2,-\i/2) -- (25.2/2,-\i/2);
}

\foreach \i in {1,...,9} {
  \draw [fill=white,very thick,color=white](2.1/2,-\i/2) +(-.25,-.21) rectangle ++(.25,.29);
  \draw(2.1/2,-\i/2) node{$x_\i$};
}

\foreach \x / \y / \n in { 25.2/1/1, 25.2/2/2, 25.2/3/3, 25.2/6/6} {
  \draw [fill=white,very thick,color=white](\x/2,-\y/2) +(-.25,-.21) rectangle ++(.25,.29);
  \draw(\x/2,-\y/2) node{ \textcolor{mymidblue}{[\n]}};
}
\foreach \x / \y / \n in { 23.4/2/19, 23.4/3/20, 23.4/7/24, 21.3/7/29} {
  \draw [fill=white,very thick,color=white](\x/2,-\y/2) +(-.27,-.21) rectangle ++(.27,.29);
  \draw(\x/2,-\y/2) node{ \textcolor{mymidblue}{[\n]}};
}
\foreach \x / \y / \n in { 25.2/4/4, 25.2/5/5, 25.2/7/7, 25.2/8/8, 25.2/9/9,
  23.4/4/21, 23.4/5/22, 23.4/6/23, 23.4/8/25, 23.4/9/26, 
  21.3/4/34, 21.3/6/35, 21.3/9/30,
  16.8/9/32,
  15/4/13, 
  15/8/17, 15/9/18,
  13.2/8/49, 13.2/9/50 } {
  \draw [fill=white,very thick,color=white](\x/2,-\y/2) +(-.25,-.21) rectangle ++(.25,.29);
  \draw(\x/2,-\y/2) node{\textcolor{mymidred}{$\overline{\n}$}};
}
\foreach \x / \y / \n in { 21.3/3/27, 21.3/5/28, 18.6/5/33, 16.8/5/31,
  15/1/10, 15/2/11, 15/3/12, 15/5/14, 15/6/15, 15/7/16,
  13.2/6/47, 13.2/7/48,
  11.1/7/53, 9.3/7/51, 9.3/9/52,
  7.5/2/40, 7.5/3/41, 7.5/5/42, 7.5/6/43, 7.5/7/44, 7.5/8/45, 7.5/9/46,
  5.7/8/56,
  4.8/1/36, 4.8/2/37, 4.8/3/38, 4.8/4/39,
  3.9/8/54, 4.8/9/55} {
  \draw [fill=white,very thick,color=white](\x/2,-\y/2) +(-.25,-.21) rectangle ++(.25,.29);
  \draw(\x/2,-\y/2) node{\n};
}

\foreach \x / \y / \z in {24.3/2/3, 24.3/4/5, 24.3/6/7, 24.3/8/9,
  22.2/3/5, 22.5/4/6,
  22.35/7/9,
  19.5/2/6, 19.8/3/7/, 20.1/4/8, 20.4/5/9,
  17.7/1/5, 15.9/5/9,
  11/2/3, 14.1/6/7, 14.1/8/9,
  12.0/6/8,12.3/7/9,
  10.2/5/7, 8.4/7/9, 4.8/5/6, 6.6/8/9,
  6.0/1/3, 6.3/2/4,
  4.8/7/8,
  3.45/1/2, 3.45/3/4, 3.0/8/9} {
  \draw [very thick] (\x/2, -\y/2) -- (\x/2,-\z/2); 
}

\foreach \i in {1,...,2} {
  \draw [very thick](-3.2/2,-\i/2) -- (-0.5/2,-\i/2);
}

\foreach \i in {1,...,2} {
  \draw [fill=white,very thick,color=white](-3.2/2,-\i/2) +(-.25,-.21) rectangle ++(.25,.29);
  \draw(-3.2/2,-\i/2) node{$x_\i$};
}
\foreach \x / \y / \n in { -0.5/1/1, -0.5/2/2} {
  \draw [fill=white,very thick,color=white](\x/2,-\y/2) +(-.25,-.21) rectangle ++(.25,.29);
  \draw(\x/2,-\y/2) node{\n};
}
\foreach \x / \y / \z in {-1.85/1/2} {
  \draw [very thick] (\x/2, -\y/2) -- (\x/2,-\z/2); 
}

\end{tikzpicture}
	\caption{Sorting networks for $n=2$ (left) and $n=9$ (right).  The main variables ${x_i}$ are on the left of the networks.  Each auxiliary variable $a_i$ is represented as integer $i$: the variables 1 to $n$ on the right are the sorted, non-increasing list; other integers are assigned sequentially by the code.  For $n=9$, the partial assignment $a_{r+1}=0$ has been applied to impose the $\leqslant_r$ constraint with $r=3$.  Known false variables are shown in red with an overbar; irrelevant variables are shown in blue with brackets.}
	\label{fig:abio2-9}
	\end{figure}
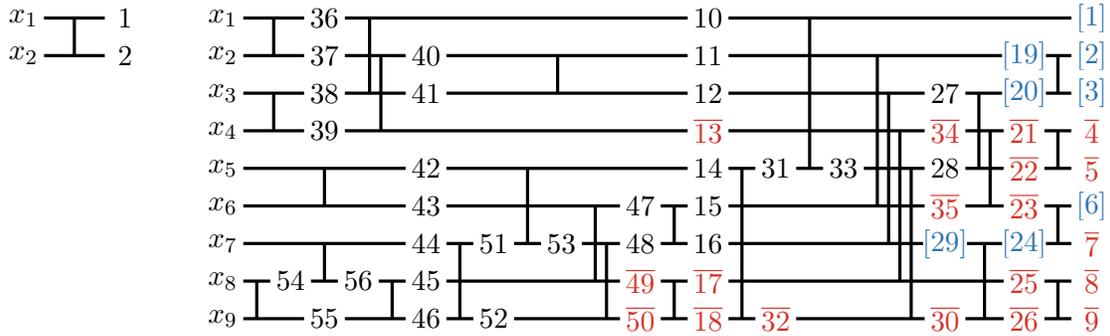

Various encodings have been presented that use sorting networks to enforce cardinal constraints.  One example is by E\'en and S\"orensson~\cite{Een:2006}, where several pseudo-boolean constraints are considered.  Codish and Zazon-Ivry~\cite{Codish:2010} suggest that pairwise sorting networks are preferable; they add extra clauses to enhance propagation -- presumably a form of strengthening.  However, Ab\'io and coworkers~\cite{Abio:2013} consider the $\leqslant_r$ constraint specifically, with detailed formulas for general $n$, and so their approach is used here as the starting-point.  The principle is that a recursive mergesort can be encoded in Conjunctive Normal Form: starting from the main variables, $(x_1,\ldots,x_n)$, the clauses produce \emph{sorted auxiliary variables}, $(a_1,\ldots,a_n)$, with the same number of 1s as the main variables but in non-increasing order.  Imposing the $\leqslant_r$ constraint is then simply achieved by setting $a_{r+1}=0$.  There is an immediate variant to the method: $a_i=0$ can also be applied for all $i>r+1$.  These variants will be called \emph{partial assignment} and \emph{full assignment} respectively.  If partial assignment, $a_{r+1}=0$, is imposed, then the sorting network guarantees that $\{a_i: i>r+1\}$ will eventually also be false, but these values are not necessarily enforced by unit propagation when the clauses are applied to a partial solution.  An example is shown in Figure~\ref{fig:abio2-9} and discussed further below.

The concept of partial and full assignment also applies if the sorting network is used to apply an equality constraint, $=_r$.  A partial assignment achieves this by imposing $a_r=1$ and $a_{r+1}=0$, whereas a full assignment imposes $a_i=1: 1\leqslant i \leqslant r$ and $a_i=0: r< i \leqslant n$.

The smallest sort, $n=2$, is shown in Figure~\ref{fig:abio2-9}, and all larger sorts are composed of multiple pairwise sorts.  The \emph{inputs}, $x_1$ and $x_2$, on the left, may be in any order, but the \emph{outputs}, $a_1$ and $a_2$, on the right, are non-increasing from top to bottom.  We will refer to the \emph{top} and \emph{bottom} outputs of a pairwise sort.  The pairwise sort can be achieved using three clauses:
\begin{equation}  \label{eq:pairforward}
\begin{gathered}
\lnot x_1 \lor a_1 \\
\lnot x_2 \lor a_1 \\
\lnot x_1 \lor \lnot x_2 \lor a_2
\end{gathered}
\end{equation}
but this is a \emph{one-way} sort only: true inputs require true outputs, but false inputs allow true or false outputs.  Conversely, if the outputs are forced by other clauses to be false, then this constrains the inputs; but if an output is true, the clauses that involve it in Eq.~\ref{eq:pairforward} are satisfied, so there is no effect on the inputs.  We will say that true values propagate to the right, and false values propagate to the left, relying on the orientation of Figure~\ref{fig:abio2-9}.  Ab\'io and coworkers~\cite{Abio:2013} note that only these clauses are required (in addition to the full or partial assignment) for $\leqslant_r$ constraints.  For $\geqslant_r$ constraints, only the reverse clauses are required:
\begin{equation}  \label{eq:pairbackward}
\begin{gathered}
x_1 \lor \lnot a_2 \\
x_2 \lor \lnot a_2 \\
x_1 \lor x_2 \lor \lnot a_1
\end{gathered}
\end{equation}
and for $=_r$ constraints, the \emph{two-way} combination of both is required.  In the context of $\leqslant_r$ constraints, \emph{one-way} will always refer to Eq.~\ref{eq:pairforward}.  The one-way clauses with partial assignment will be called the \emph{basic} or \emph{unstrengthened} encoding of this type, and the two-way clauses with full assignment will be called \emph{fully strengthened}.  Section~\ref{sec:compare} investigates the application of two-way sorts in $\leqslant_r$ constraints as a form of strengthening: are the extra clauses in Eq.~\ref{eq:pairbackward} justified by faster solution?

Ab\'io and coworkers present ways to simplify the sorting network when only some of the outputs of a sorting network are of interest -- for example, when only $a_{r+1}$ or $\{a_{i}: i>r\}$ are required.  However, a different approach is developed here: the full network is considered, but some variables are noted to be `irrelevant', `known true' or `known false'.  Variables with any of these statuses do not need to be included in clauses; other variables will be called `active'.  If a variable is irrelevant, in that a true or false value will not affect the active variables, then clauses that involve it can also be omitted.  An example of this, $n=9$, $r=3$ with partial assignment, is shown in Figure~\ref{fig:abio2-9}.  The partial assignment defines $a_4$ to be `known false', by definition.  Other variables can be deduced to be `known false'.  If only one-way equations are applied, then `known false' status can propagate left (in particular from the top output of a pairwise sort), for example from $\lnot a_4$ to $\lnot a_{21}$ and $\lnot a_{22}$.  However, the two-way nature of the sorting network can be used to propagate `known false' status to the right: from  $\lnot a_{21}$ and $\lnot a_{22}$ to $\lnot a_5$, and also from $\lnot a_{18}$ to $\lnot a_{32}$ etc.

	\begin{figure}[ht] 
	\centering

\begin{tikzpicture}
\definecolor{mygrey}{rgb}{0.7,0.7,0.7}
\definecolor{mymidblue}{rgb}{0.1718,0.4531,0.7031}
\definecolor{mydarkblue}{rgb}{0.0859,0.2265,0.3515}
\definecolor{mymidred}{rgb}{0.8359,0.1875,0.1484}
\definecolor{mydarkred}{rgb}{0.4179,0.0938,0.0742}

\foreach \i in {1,...,4} {
  \draw [very thick](-3.2/2,-\i/2) -- (4.9/2,-\i/2);
}

\foreach \i in {1,...,4} {
  \draw [fill=white,very thick,color=white](-3.2/2,-\i/2) +(-.25,-.21) rectangle ++(.25,.29);
  \draw(-3.2/2,-\i/2) node{$x_\i$};
}
\foreach \x / \y / \n in { -0.5/1/5, -0.5/2/6, -0.5/3/7, -0.5/4/8, 2.2/2/9, 2.2/3/10} {
  \draw [fill=white,very thick,color=white](\x/2,-\y/2) +(-.25,-.21) rectangle ++(.25,.29);
  \draw(\x/2,-\y/2) node{\n};
}
\foreach \x / \y / \n in { 4.9/1/1, 4.9/2/2, 4.9/4/4} {
  \draw [fill=white,very thick,color=white](\x/2,-\y/2) +(-.25,-.21) rectangle ++(.25,.29);
  \draw(\x/2,-\y/2) node{ \textcolor{mymidblue}{[\n]}};
}
\foreach \x / \y / \n in { 4.9/3/3 } {
  \draw [fill=white,very thick,color=white](\x/2,-\y/2) +(-.25,-.21) rectangle ++(.25,.29);
  \draw(\x/2,-\y/2) node{\textcolor{mymidred}{$\overline{\n}$}};
}
\foreach \x / \y / \z in {-1.85/1/2, -1.85/3/4, 0.7/1/3, 1.0/2/4, 3.55/2/3} {
  \draw [very thick] (\x/2, -\y/2) -- (\x/2,-\z/2); 
}


\foreach \i in {1,...,2} {
  \draw [very thick](7.6/2,-\i/2) -- (10.3/2,-\i/2);
}

\foreach \x / \y / \n in { 7.6/1/6, 7.6/2/8, 10.3/1/9, 10.3/2/2} {
  \draw [fill=white,very thick,color=white](\x/2,-\y/2) +(-.25,-.21) rectangle ++(.25,.29);
  \draw(\x/2,-\y/2) node{\n};
}
\foreach \x / \y / \n in { 10.3/2/4} {
  \draw [fill=white,very thick,color=white](\x/2,-\y/2) +(-.25,-.21) rectangle ++(.25,.29);
  \draw(\x/2,-\y/2) node{ \textcolor{mymidblue}{[\n]}};
}
\foreach \x / \y / \z in {8.95/1/2} {
  \draw [very thick] (\x/2, -\y/2) -- (\x/2,-\z/2); 
}

\foreach \i in {1,...,2} {
  \draw [very thick](-8.6/2,-\i/2) -- (-5.9/2,-\i/2);
}

\foreach \x / \y / \n in { -8.6/1/14,  -5.9/1/31} {
  \draw [fill=white,very thick,color=white](\x/2,-\y/2) +(-.25,-.21) rectangle ++(.25,.29);
  \draw(\x/2,-\y/2) node{\n};
}
\foreach \x / \y / \n in { -8.6/2/18, -5.9/2/32} {
  \draw [fill=white,very thick,color=white](\x/2,-\y/2) +(-.25,-.21) rectangle ++(.25,.29);
  \draw(\x/2,-\y/2) node{\textcolor{mymidred}{$\overline{\n}$}};
}
\foreach \x / \y / \z in {-7.25/1/2} {
  \draw [very thick] (\x/2, -\y/2) -- (\x/2,-\z/2); 
}

\end{tikzpicture}
	\caption{From left to right: A pairwise sort from the $n=9$, $r=3$ network of Figure~\ref{fig:abio2-9};  Sorting network for $n=4$ with partial assignment of $r=3$;  A pairwise sort from this network.  Symbols are as in Figure~\ref{fig:abio2-9}.}
	\label{fig:abio4}
	\end{figure}
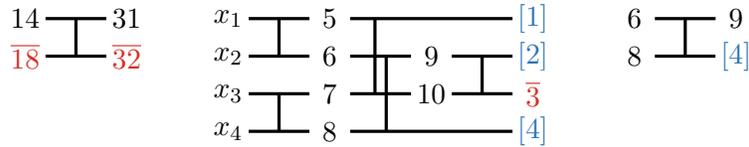 

When the statuses of all variables have been calculated, there are sometimes opportunities for removing `active' variables, as well as the known and irrelevant ones.  This is illustrated in Figure~\ref{fig:abio4}.  In a pairwise sort, if the statuses of one input and the bottom output are both `known false', then the remaining input and top output are logically equivalent. (At least, this is true in the presence of two-way pair comparison.  For one-way sorting, it is easiest to assume a two-way strengthening for that pairwise sort.)  Therefore, a single auxiliary variable can replace both of them.  This saves one variable and two clauses, with no change in the solution logic.  An example is shown in the left of Figure~\ref{fig:abio4} (and a second can be found in the right of Figure~\ref{fig:abio2-9}).

A more specialised vertex removal is possible if, with one-way pairwise sorts, the bottom output of a pairwise sort has status `irrelevant' and the other three variables are `active'.  (An example is shown in the right of Figure~\ref{fig:abio4}.)  In this case, the three active variables can be replaced by single variable.  To show this, we consider solutions in the auxiliary variables that are consistent with the one-way sorting clauses, Eq.~\ref{eq:pairforward}, and an ordered set of main-variable values.  If this set of main-variable values is consistent with the $\leqslant_r$ constraint, then all auxiliary-variable solutions will obey $a_{r+1}=0$; if it breaks the constraint, all solutions will violate $a_{r+1}=0$.  This is a consequence of the sorting network acting as it must.  We will show that all solutions can be adjusted so that the three variables all have the same value, while staying consistent with the relevant sorting clauses and not changing the sorted auxiliary variables.  If the top output is false in any solution, then both inputs are false, by Eq.~\ref{eq:pairforward}.  So, the only solutions that require any adjustment have the top output true.  In that case, the relevant clauses of Eq.~\ref{eq:pairforward} are satisfied by the top output, and the inputs can be changed to true (if necessary) and these clauses will remain satisfied.  The inputs will be involved in other pairwise sorts to their left, but there they will act as outputs, so changing them to true will not invalidate any clause in Eq.~\ref{eq:pairforward}.  So, every ordered set of main-variable values is consistent with values for the auxiliary variables where the three variables are equal or, equivalently, replaced by a single variable.

To put this in perspective, the first kind of vertex removal reduces the number of auxiliary variables by an average of approximately 0.2\% in the test cases of Section~\ref{sec:compare}.  The second kind reduces it by an average of approximately 10\% for the partial one-way variant, which is the only variant where it applies.  The average of 10\% is skewed by some large savings for small cases.  For larger cases, the average is approximately 7\%.

\subsection{Comparison of encodings in theory}
\label{sec:comparetheory}

The previous sections have developed variants of the published encodings.  These are applied in practice in the next section, but some initial comparisons are made here.

One check on the correctness of a cardinality encoding is that it should allow all permutations of the main variables that obey the cardinality constraint, and no others.  (A convenient way to perform this check is to use the {\sc picosat} solver~\cite{picosat:web}, with its option \texttt{--all}.  This option is recommended for all \Sat solvers whenever possible.)  So, for example, suppose $n=10$ and $r=4$: an exhaustive list of solutions for $\leqslant_r$ should comprise $\left( \binom{10}{0}+\binom{10}{1}+\binom{10}{2}+\binom{10}{3}+\binom{10}{4} \right)=386$ solutions in the main variables; an exhaustive list for $=_r$ should comprise $\binom{10}{4}=210$ main-variable solutions.  Naturally, the encodings considered here all pass this check.  It is interesting, however, to note in Table~\ref{tab:multiples} the number of solutions for each encoding, especially in unstrengthened or partially strengthened variants.  Any excess over 386 or 210 indicates multiple solutions in the auxiliary variables.  It can be seen that the numbers of multiple solutions are sometimes large, even for small values of $n$ and $r$.  The tests in Table~\ref{tab:multiples} took several hours on the computer described in Section~\ref{sec:methodology}.

\begin{table}[htbp]
\centering
\caption{Numbers of solutions for encodings with $n=10$ and $r=4$.}
\begin{tabular}{l r}
\hline
{\bf Sequential-counter encoding:}\\
Unstrengthened  & 10371 \\
Strengthened with Eq.~\ref{eq:sinz3} & 3360 \\
Strengthened with Eq.~\ref{eq:sinz4} & 888 \\
Fully strengthened with both equations & 386 \\
Equality constraint & 210 \\
\hline
{\bf Tree-based encoding:}\\
Unstrengthened (Figure~\ref{fig:bail}) & 8474 \\
With sideways strengthening & 5120 \\
With inequality strengthening & 1646 \\
With both strengthenings & 1645 \\
Equality constraint (Figure~\ref{fig:bailexact}) & 210 \\
\hline
{\bf Sort-based encoding:}\\
Partial assignment, one-way clauses & 1115475 \\
Full assignment, one-way clauses & 180770 \\
Partial assignment, two-way clauses & 386 \\
Full assignment, two-way clauses & 386 \\
Full assignment of equality constraint & 210 \\
\hline
\end{tabular}
\label{tab:multiples}
\end{table}

\begin{table}[htbp]
\centering
\caption{Numbers of auxiliary variables, clauses, literals and main-variable literals (MVLs) in the encodings for $\leqslant_r$ or $=_r$ with $n=66$, $r=36$.  These constraints are used in Section~\ref{sec:compare} for sequence A227116(11).  To set up that problem, 315 clauses (each of size 3, hence using 945 literals) are required in addition to the constraint's clauses.}
\begin{tabular}{l r r r r}
\cline{2-5}
& Variables & Clauses & Literals & MVLs \\
\hline
{\bf Sequential-counter encoding:}\\
Unstrengthened  & 1080 & 2154 & 5358 & 1110 \\
Equality constraint & 1080 & 4320 & 10734 & 2226 \\
\hline
{\bf Tree-based encoding:}\\
Unstrengthened (Figure~\ref{fig:bail}) & 328 & 1402 & 3854 & 132 \\
Equality constraint (Figure~\ref{fig:bailexact}) & 328 & 3080 & 8254 & 264\\
\hline
{\bf Sort-based encoding:}\\
Partial assignment, one-way clauses & 846 & 1296 & 3047 & 132\\
Full assignment of equality constraint & 904 & 2778 & 6460 & 264 \\
\hline
\end{tabular}
\label{tab:counts}
\end{table}

Bailleux and Boufkhad~\cite{Bailleux:2003} state that their encoding requires $O(n \log(n))$ variables and $O(n^2)$ clauses, and these limits have been quoted in other publications~\cite{Frisch:2010}\cite{Sinz:2005}.  However, it should be recalled that these are upper limits, and it should not be concluded that the tree-based encoding is excessively large, for example when $r$ or $(n-r)$ is small.  In those cases, the pseudocode of Figure~\ref{fig:bail}, at least, produces small clause counts.  The variable count is $O(r n)$, specifically $\leqslant r(n-2)$, as well as $O(n \log(n))$.  More concretely, the tree-based encoding in Figure~\ref{fig:bail} produces the lowest clause count (or, occasionally, equal-lowest) of the three unstrengthened encodings for all combinations $(n,r)$ satisfying $1\leqslant r<n<1000$.  This contradicts a claim made by Sinz~\cite{Sinz:2005} that the sequential-counter encoding ``performs better [with respect to number of clauses required] for small values of'' $r$.  This may be because Figure~\ref{fig:bail} is more efficient than previous encodings.  However, one of the firmest conclusions from the next section is that clause counts, and other measures of encoding size, are \emph{not} a reliable indicator of how well encodings perform in practical use.

Encoding sizes for the different methods and variants are shown in Table~\ref{tab:counts} for a specific example.  The sequential-counter encoding has the largest encoding size by any of the measures shown.  For each method, the equality constraint requires approximately twice as many clauses as the inequality constraint with one-way clauses, but effectively the same number of variables.  The tree-based encoding has a substantially smaller number of auxiliary variables than the other methods.  The number of clauses is not so much smaller, indicating that the auxiliary variables are more inter-related for the tree-based encoding.  The main variables are involved much more frequently in the sequential-counter encoding than in the others.  All these observations typically apply to the constraints used in Section~\ref{sec:compare}; further examples can be seen in Figure~\ref{fig:orderings}.  The overall encoding sizes in that section are dominated by the constraint encodings.

\section{Comparison of encodings in practice}
\label{sec:compare}

\subsection{Test cases}
\label{sec:testcases}

Five test cases are considered, and are referred to by their OEIS sequence numbers~\cite{oeis:web}; see Table~\ref{tab:oeis}.  An example is A240443, which can be defined as follows: Consider an $L \times L$ square grid of points. What is the minimum number of points that can be selected, such that every square of points contains at least one of the selection?  This number is A240443$(L)$.  If only grid-aligned squares are considered, then the minimum number of points is A152125$(L)$.  The other examples use an equilateral triangular grid, with $L$ points along each edge, and similarly look for the minimum selection to be in every one of a set of equilateral triangles.  The sequences A319158, A227116 and A319159 differ only in the kinds of triangles considered: aligned with the grid and pointing in the same direction; sides parallel to the grid, including upside-down; and any orientation, respectively.

These test cases are easily stated in Conjunctive Normal Form: a main variable is assigned to each point (where true indicates selected), and a clause is formed of the points in each square or triangle.  These clauses use the main variables only in positive polarity.  A cardinality constraint, $\leqslant_r$ or $=_r$ is then applied, with $n=L^2$ for squares or $n=L(L+1)/2$ for triangles.  The appropriate cardinality limit, $r$, must be found by experimentation.  For a sequence $a(L)$, then $r=a(L)$ is the minimal satisfiable constraint (abbreviated as SAT), and $r=a(L)-1$ is the maximal unsatisfiable constraint (UNSAT).

\begin{table}[htbp]
\centering
\caption{Values of the sequences used in test cases.}
\begin{tabular}{|l| c c c c c c c c c c c c c c|}
\hline
$L$ & 2 & 3 & 4 & 5 & 6 & 7 & 8 & 9 & 10 & 11 & 12 & 13 & 14 & 15\\
\hline
A152125$(L)$ & 1 & 2 & 4 & 8 & 12 & 17 & 23 & 30 & 39 & & & & & \\
A240443$(L)$ & 1 & 3 & 6 & 10 & 15 & 21 & 27 & 34 & 42 & & & & & \\
A319158$(L)$ & 1 & 2 & 4 & 6 & 9 & 13 & 18 & 23 & 29 & 35 & 43 & 51 & & \\
A227116$(L)$ & 1 & 2 & 4 & 7 & 9 & 14 & 18 & 23 & 29 & 36 & 44 & 52 & 61 & 71 \\
A319159$(L)$ & 1 & 2 & 4 & 7 & 11 & 16 & 22 & 28 & 35 & 44 & 53 & 63 & 74 & 86 \\
\hline
\end{tabular}
\label{tab:oeis}
\end{table}

\subsection{Methodology}
\label{sec:methodology}

The methodology used here is to apply Knuth's {\sc{sat13}} solver to the test cases defined in the previous section, and to compare the computational effort required to solve each case using various encodings of the cardinality constraint.  This solver uses Conflict-Driven Clause Learning; according to its author, it `decently represents the main CDCL paradigms', while not being intended as a cutting-edge competitor~\cite{Knuth:web}.  It was chosen precisely because it is stable and comprehensible, and because it calculates the number of  memory accesses or \emph{mem count}, as a measure of computational effort.  The calculated mem counts are not dependent on computer architecture or software.  Each \emph{mem} is an access to a 64-bit word by the program.

\begin{figure}[htp] 
	\centering
{\scalebox{1.0}{\includegraphics{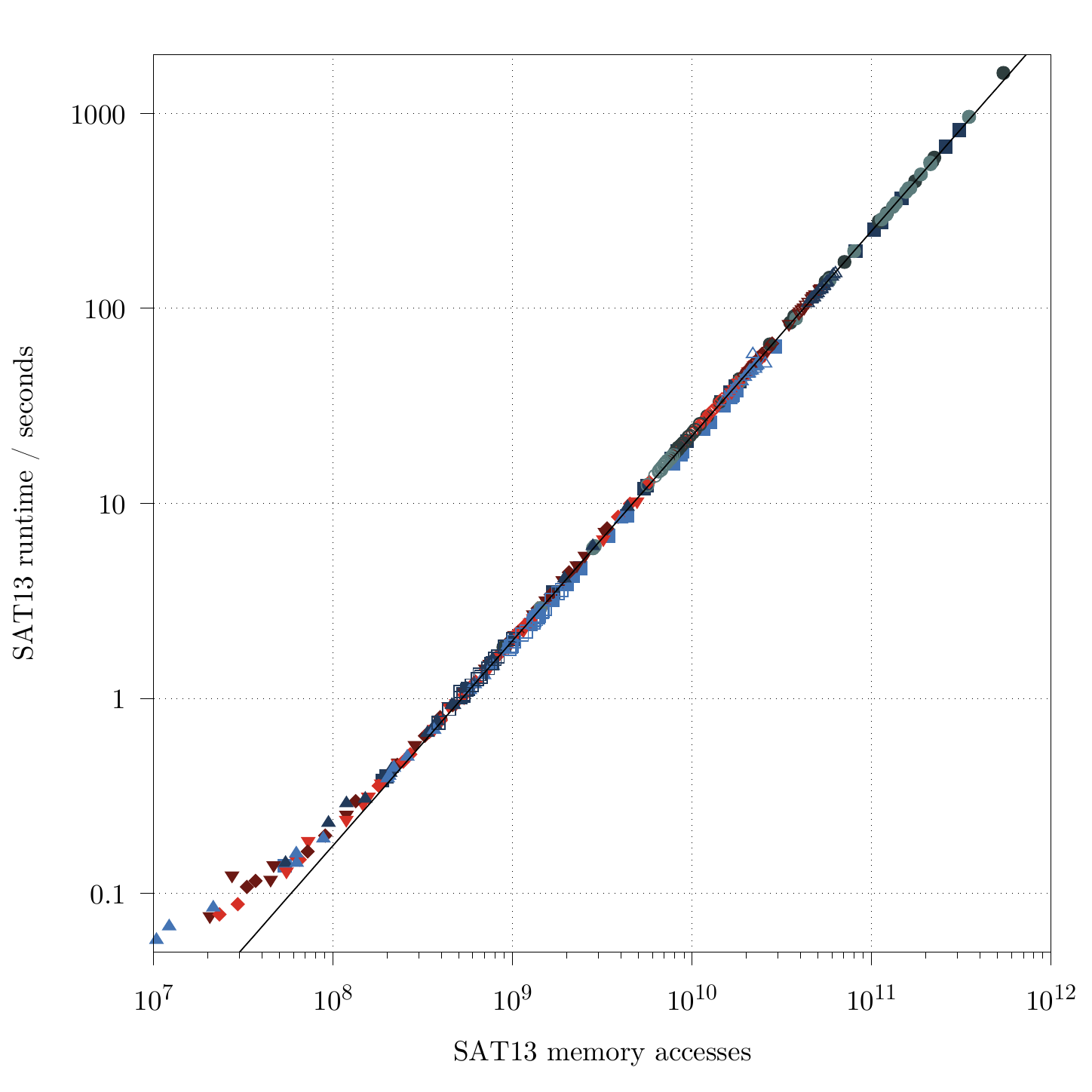}}}
	\caption{Runtime plotted against mem count for the {\sc{sat13}} solver, when applied to the `largest solved' cases in this section (as defined in the main text) with the unstrengthened sequential-counter and tree-based encodings.  The line is the least-squares best fit to all log-log data above 1~second, and has a gradient of 1.050.  Symbols are the same as in Figure~\ref{fig:sinzvariants}, except that symbols from the tree-based encoding are darker than those from the sequential-counter encoding.  A mem count of $10^{11}$ corresponds to a runtime of 250~seconds.}
	\label{fig:mems_vs_time}
	\end{figure}

\begin{figure}[htp] 
	\centering
{\scalebox{1.0}{\includegraphics{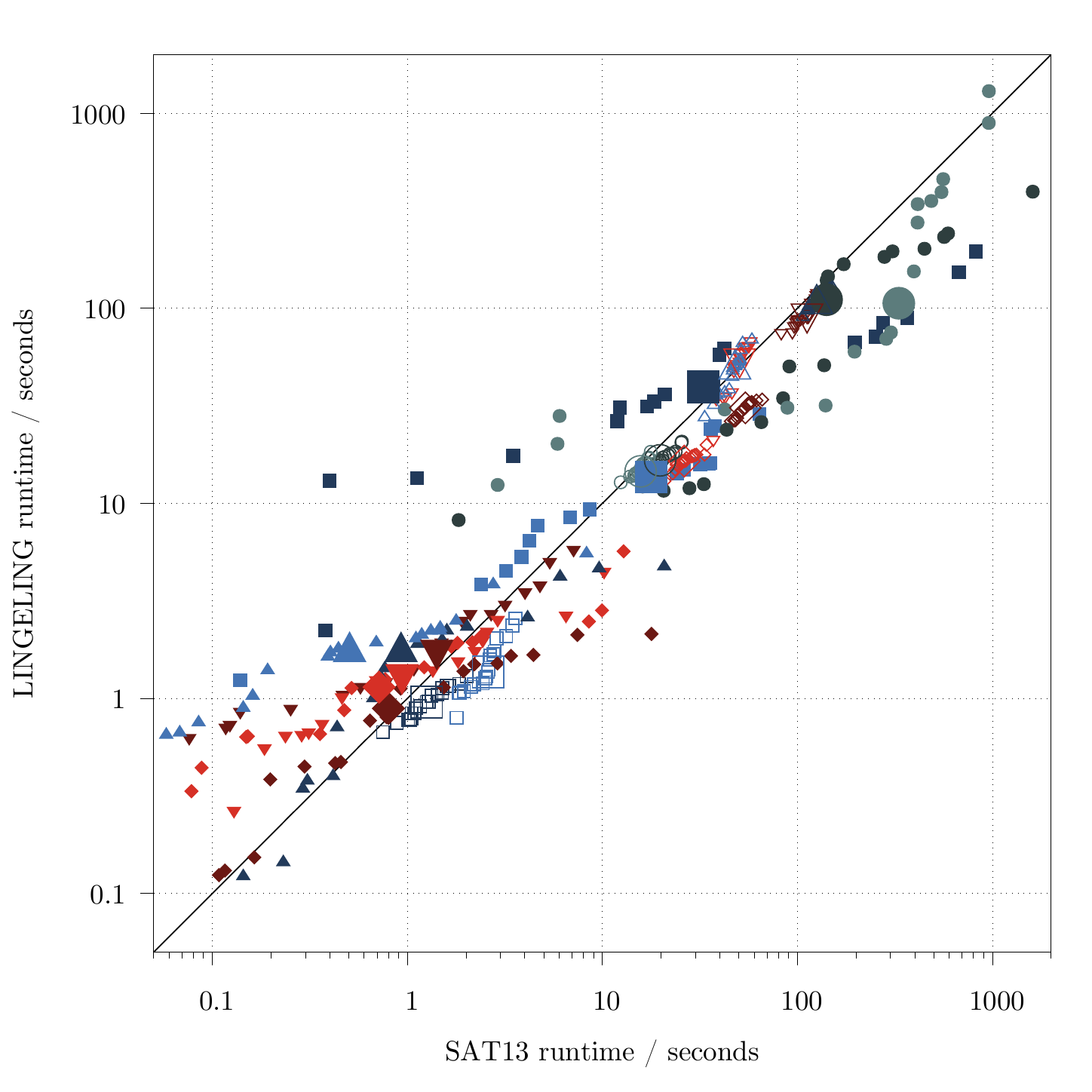}}}
	\caption{Runtime for the {\sc{lingeling}} solver plotted against runtime for the {\sc{sat13}} solver, when applied to the `largest solved' cases in this section (as defined in the main text) with the unstrengthened sequential-counter and tree-based encodings.  The line represents equality.  Each solver was applied to each case 19 times, with different random seeds; the $i$-th ordered result for one solver is plotted against the $i$-th for the other.  The median points are plotted with larger symbols.  Symbols are the same as in Figure~\ref{fig:sinzvariants}, except that symbols from the tree-based encoding are darker than those from the sequential-counter encoding.}
	\label{fig:time_vs_time}
	\end{figure}

It is reasonable to ask whether mem count is indeed related to runtime for the {\sc{sat13}} solver, and whether this in turn is related to runtime for a more sophisticated solver.  These questions are addressed in Figure~\ref{fig:mems_vs_time} and Figure~\ref{fig:time_vs_time}.  Timings here are for a computer running Windows 10~Pro operating system on a Xeon~E3-1240v5 processor (launched in 2015 and marketed for `entry-level servers and workstations'~\cite{intel:2015}).  System time was reported by the {\texttt{time}} command in the Linux Bash Shell on Windows~10, with no other user processes running.  (The importance of this last proviso makes runtime an inconvenient measure.  The runtime for a run limited to $10^{11}$ mems could be as high as 1200~seconds when other processes were running, compared to $<\!250$ seconds for an otherwise unloaded processor.  `CPU time' is not a simple concept for a modern processor.)  The {\sc{lingeling}} solver~\cite{Biere:web} is a successful one, under active development; a recent version (version~{\sc{bcj}}, released May~2018) was compared with {\sc{sat13}}.  All programs were compiled using the {\sc{gcc}} compiler, version~7.1.0, with flag {\texttt{-O3}}.  In Figure~\ref{fig:mems_vs_time} and Figure~\ref{fig:time_vs_time}, the `largest solved' cases comprise the largest size for each test case in Section~\ref{sec:testcases} where all three encodings completed with a median mem count less than $10^{11}$.  (A small exception to this is the A319159~SAT case, where size $L=12$ was run even though only 7 out of 19 repeats of the sequential-counter method's unstrengthened variant completed inside the criterion.)

Figure~\ref{fig:mems_vs_time} shows that mem count is very closely, almost linearly, related to runtime for a specific computer.  Figure~\ref{fig:time_vs_time} shows that the {\sc{sat13}} solver is a respectable option, in the same ballpark of runtimes as the {\sc{lingeling}} solver.  The {\sc{sat13}} solver appears to have more variability in its results than the {\sc{lingeling}} solver: higher maxima, but sometimes also lower minima.  The shortest runtime in the 19 repeats of the {\sc{sat13}} solver sometimes achieves a substantial victory over the corresponding shortest {\sc{lingeling}} runtime.

Variability of runtime or mem count was a significant factor.  All test cases were run with 19 repeats.  The medians of the results are used.  Error bars are plotted in Figure~\ref{fig:orderings}.  These error bars represent the dispersion of the logarithm of the mem-count, using the robust measure $S_n$ developed by Rousseeuw and Croux~\cite{rousseeuw:1993}.  If the log-data followed a Gaussian distribution, this measure would equal the standard deviation of individual samples.  The dispersions of the results for different encodings appear to be approximately equal for each test case.  The dispersion of SAT runs is considerably larger than that of UNSAT runs, and it is relatively difficult to make confident conclusions from comparisons of SAT mem counts.  Error bars are not plotted in other figures, in order to avoid clutter.  The dispersion values indicated in Figure~\ref{fig:orderings} are representative of all test cases.

The smallest problem sizes considered in this section represent very small challenges to the \Sat solvers; they would be difficult to quantify using runtimes, which are less than 1~second.  Mem count is useful here, and the small sizes are included mainly because they indicate the growth rate as size increases.  At the other extreme, the runs reported in later sections were abandoned if the mem count exceeded $10^{11}$.

\subsection{Results of test cases}
\label{sec:results}

\begin{figure}[htp] 
	\centering
{\scalebox{1.0}{\includegraphics{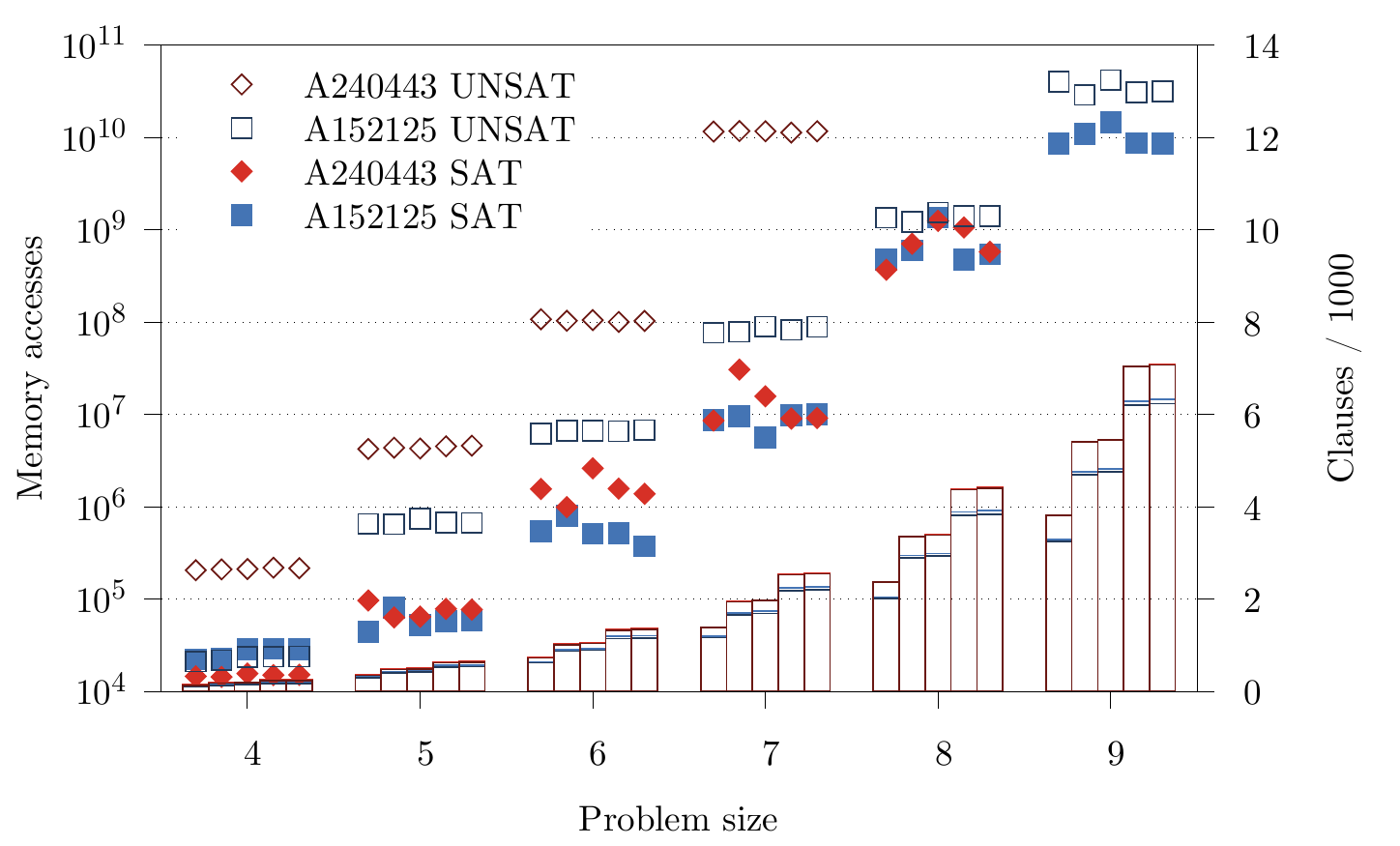}}}
{\scalebox{1.0}{\includegraphics{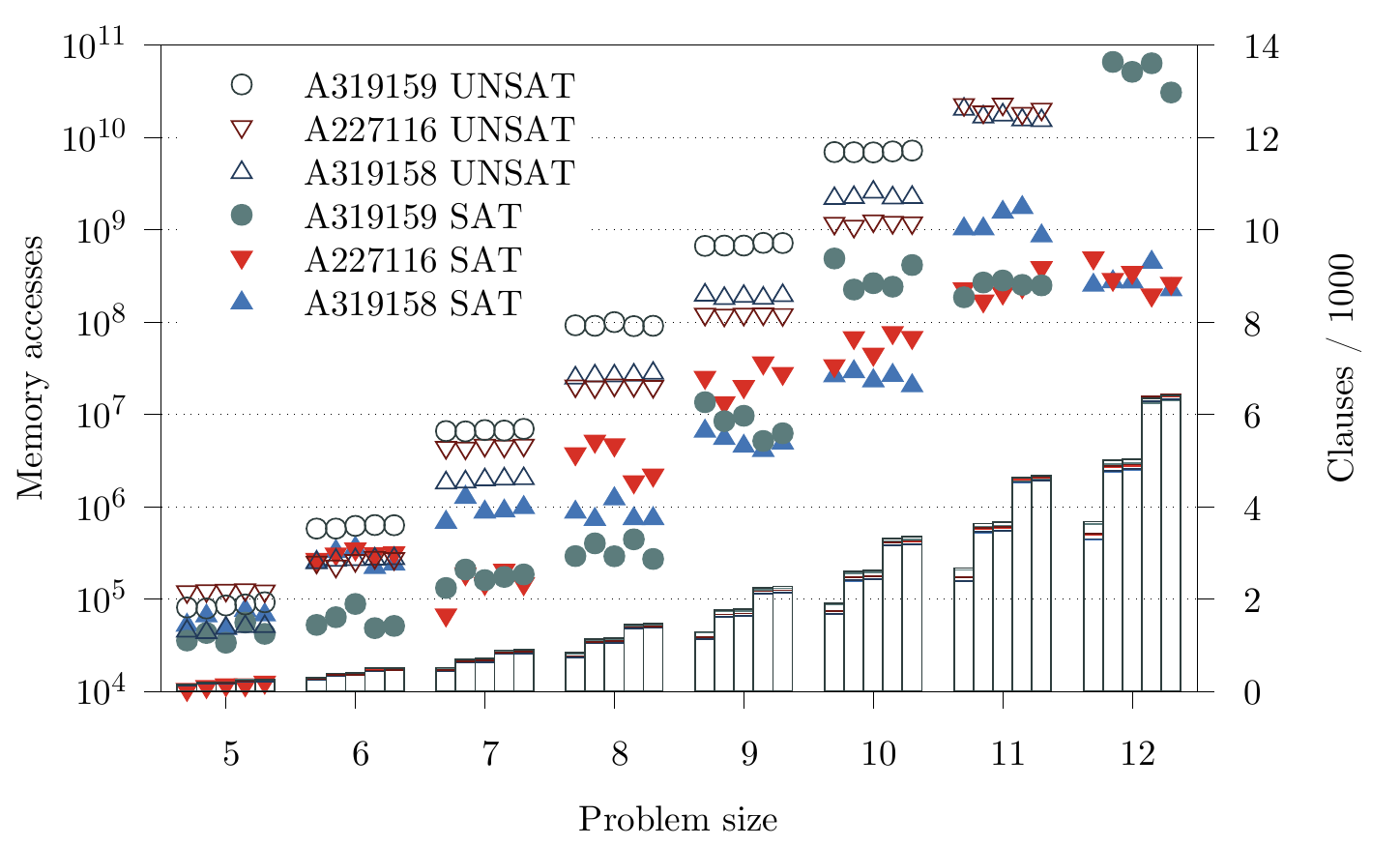}}}
	\caption{Results for variants of the sequential-counter encoding discussed in Section~\ref{sec:sinz}.  Top: test cases involving squares; bottom: test cases involving triangles.  The numbers of memory accesses are plotted as symbols, using the logarithmic scale on the left.  The numbers of clauses are plotted as bars, using the linear scale on the right. At each size, the variants are, from left to right: the unstrengthened variant (Eq.~\ref{eq:sinz1} and~\ref{eq:sinz2}); strengthened with Eq.~\ref{eq:sinz3}; strengthened with Eq.~\ref{eq:sinz4}; fully strengthened with both these; fully strengthened and converted to equality constraint.}
	\label{fig:sinzvariants}
	\end{figure}

\begin{figure}[htp] 
	\centering
{\scalebox{1.0}{\includegraphics{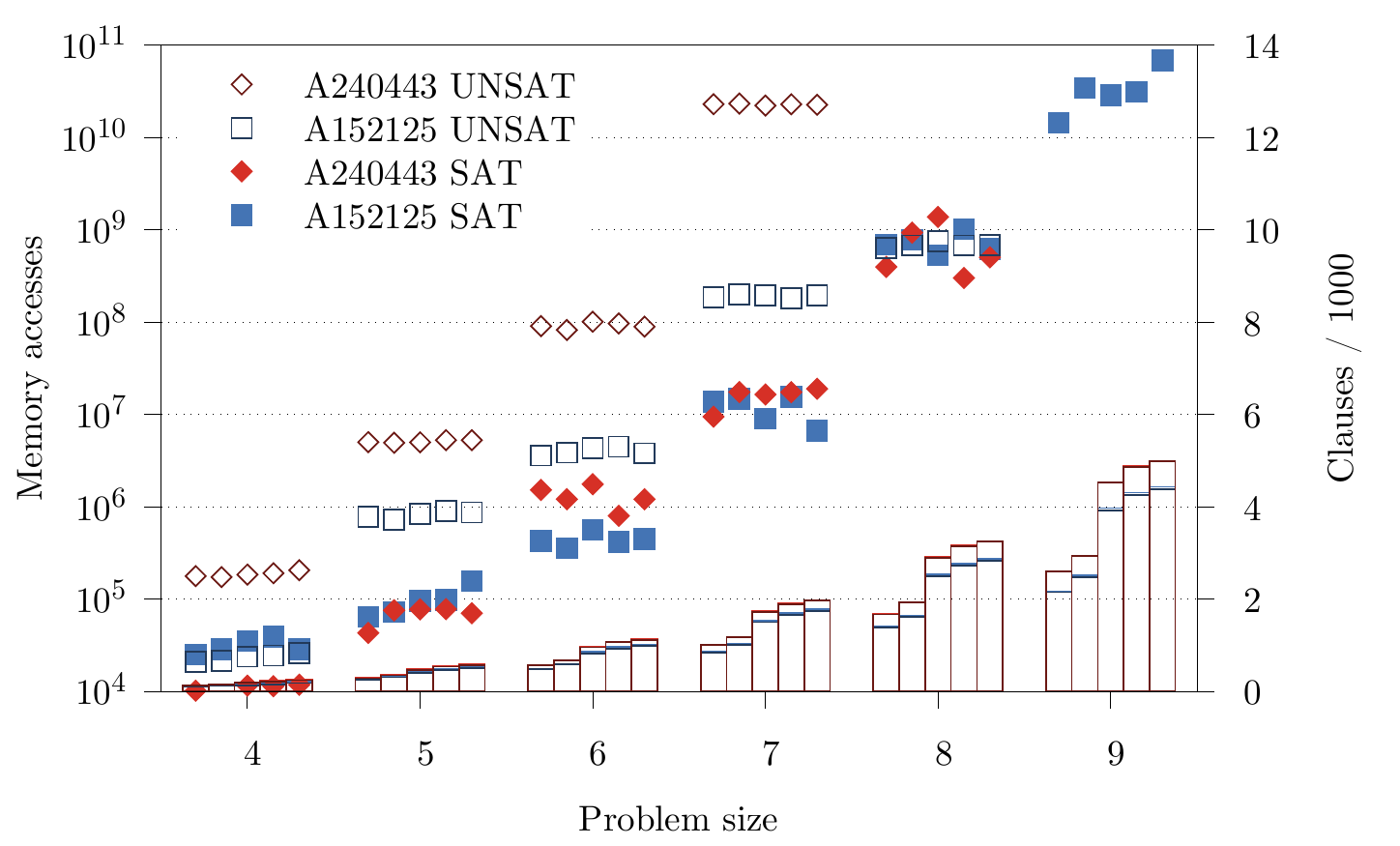}}}
{\scalebox{1.0}{\includegraphics{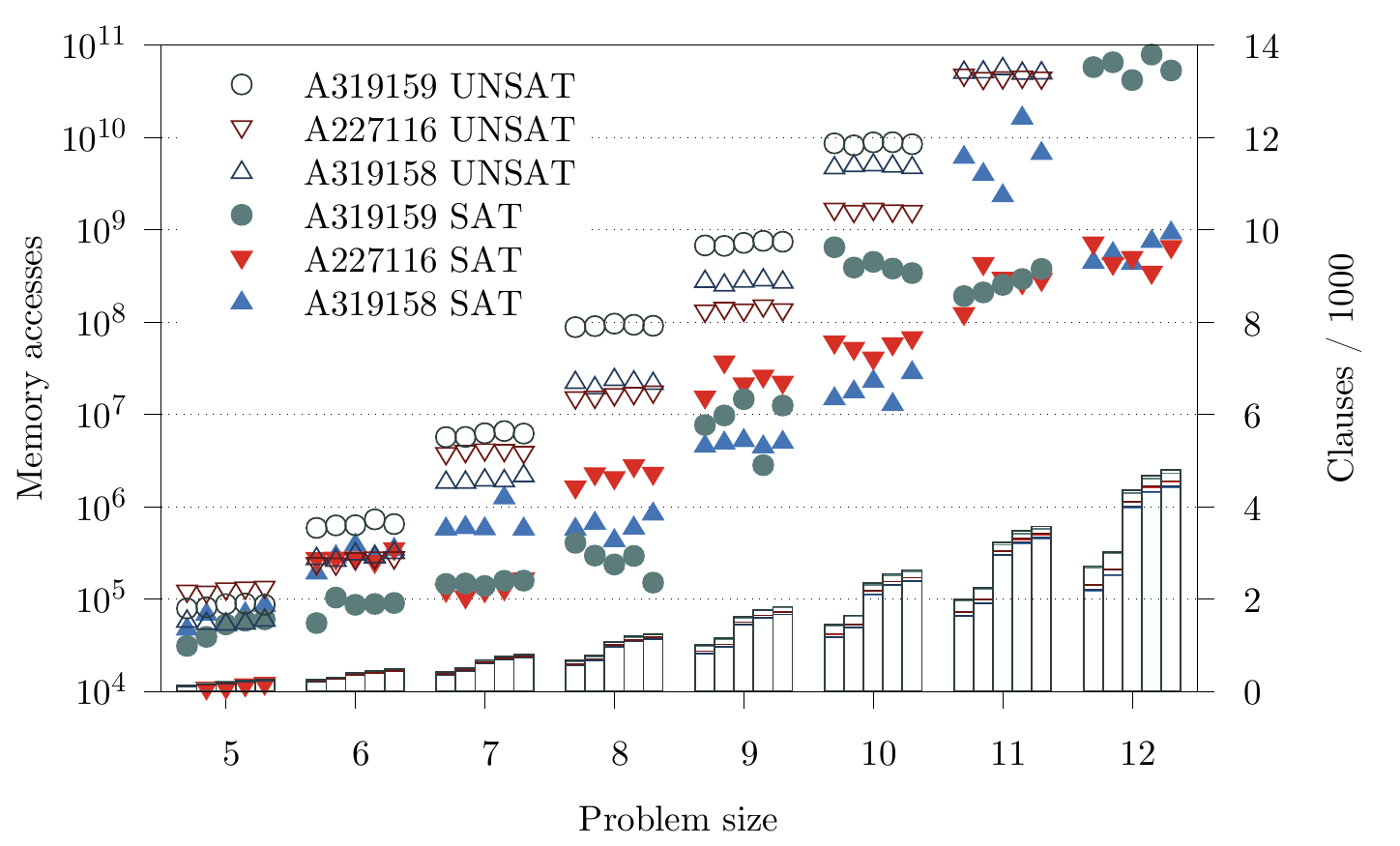}}}
	\caption{Results for variants of the tree-based encoding discussed in Section~\ref{sec:bail}.  Top: test cases involving squares; bottom: test cases involving triangles.  The numbers of memory accesses are plotted as symbols, using the logarithmic scale on the left.  The numbers of clauses are plotted as bars, using the linear scale on the right. At each size, the variants are, from left to right: the unstrengthened variant (Figure~\ref{fig:bail}); with sideways strengthening; with inequality strengthening; with both strengthenings; and converted to equality constraint (Figure~\ref{fig:bailexact}).}
	\label{fig:bailvariants}
	\end{figure}

\begin{figure}[htp] 
	\centering
{\scalebox{1.0}{\includegraphics{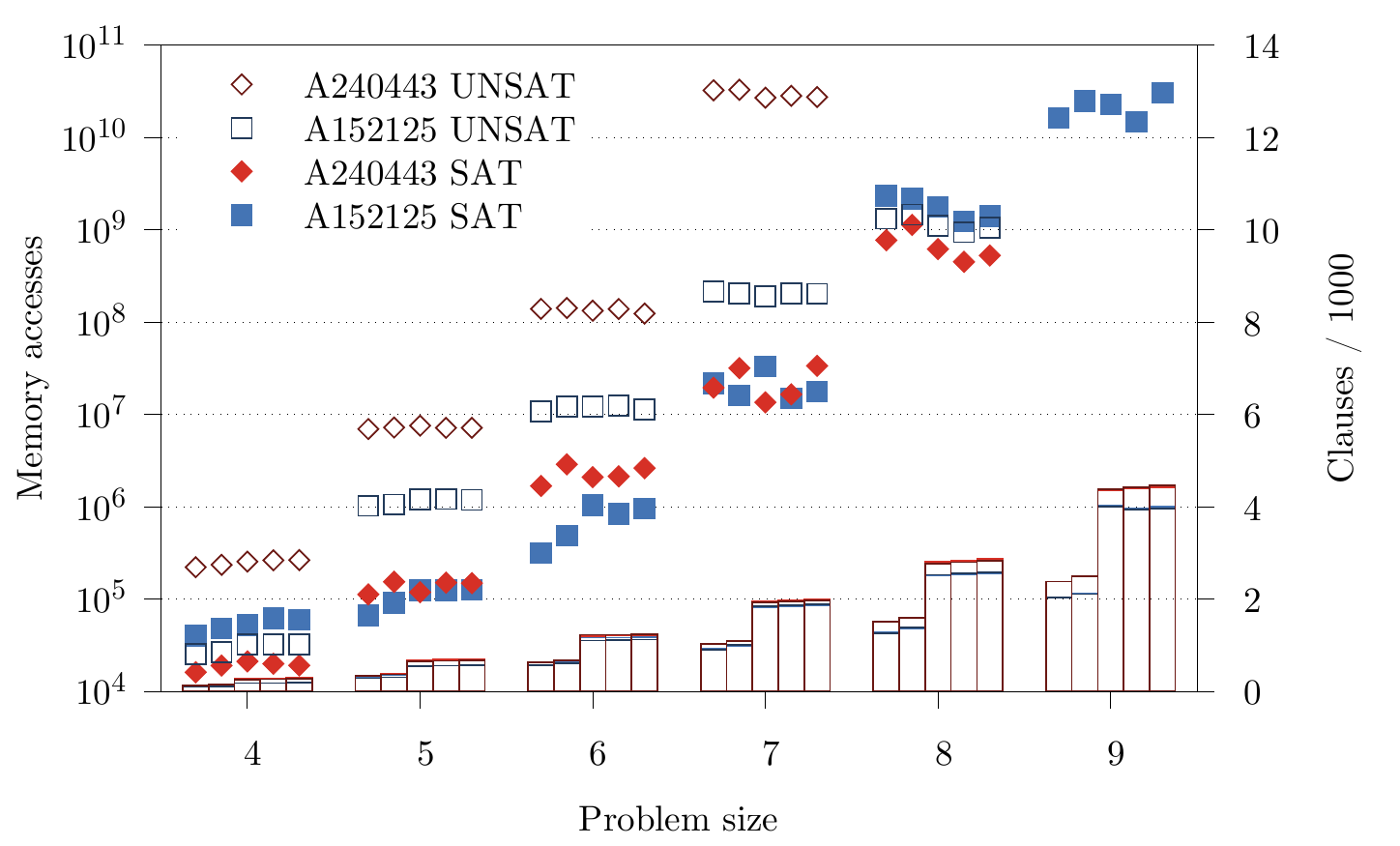}}}
{\scalebox{1.0}{\includegraphics{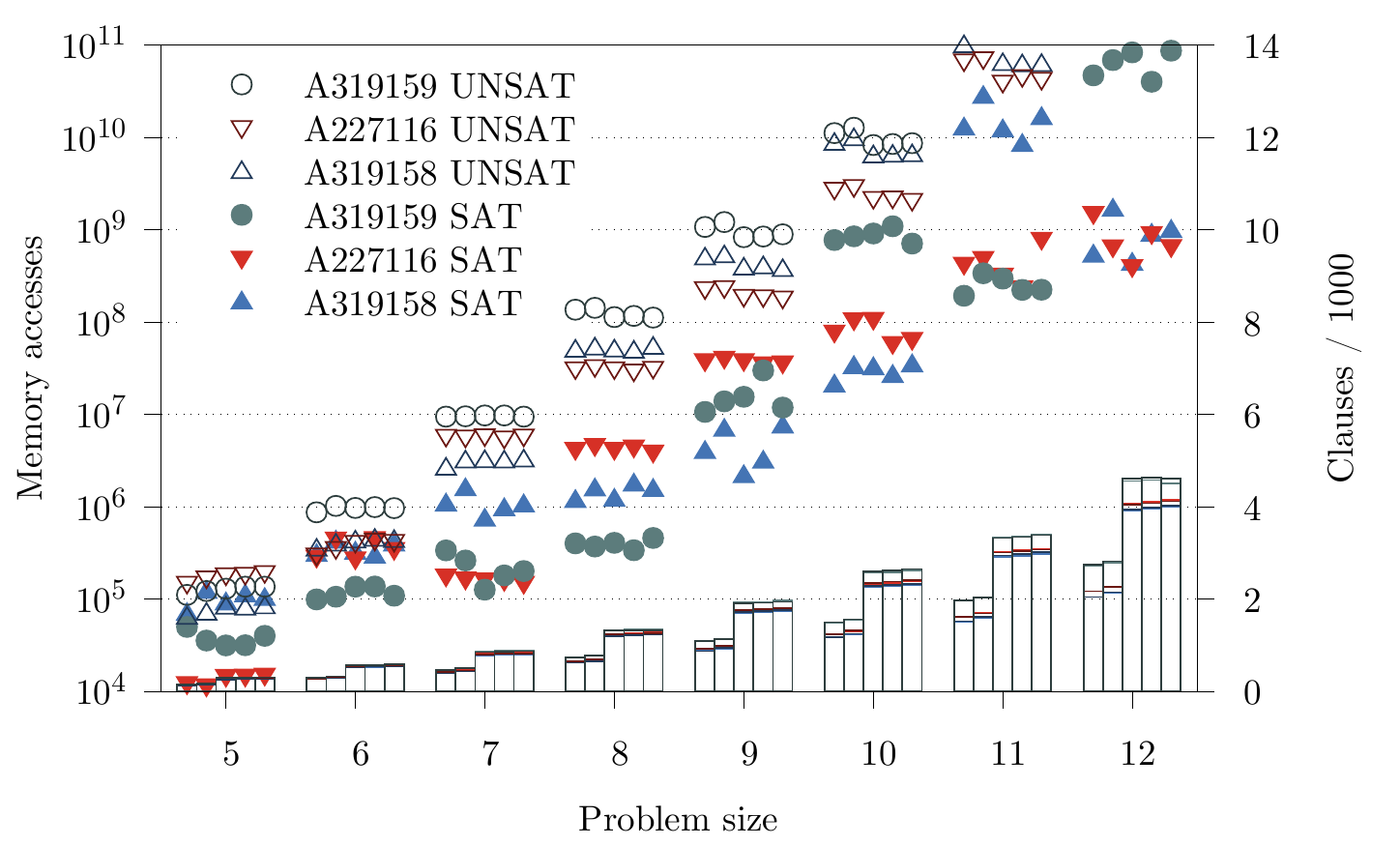}}}
	\caption{Results for variants of the sort-based encoding discussed in Section~\ref{sec:abio}.  Top: test cases involving squares; bottom: test cases involving triangles.  The numbers of memory accesses are plotted as symbols, using the logarithmic scale on the left.  The numbers of clauses are plotted as bars, using the linear scale on the right. At each size, the variants are, from left to right: partial assignment with one-way clauses; full assignment with one-way clauses; partial assignment with two-way clauses; full assignment with two-way clauses; and full assignment of equality constraint.}
	\label{fig:abiovariants}
	\end{figure}

\begin{figure}[htp] 
	\centering
{\scalebox{1.0}{\includegraphics{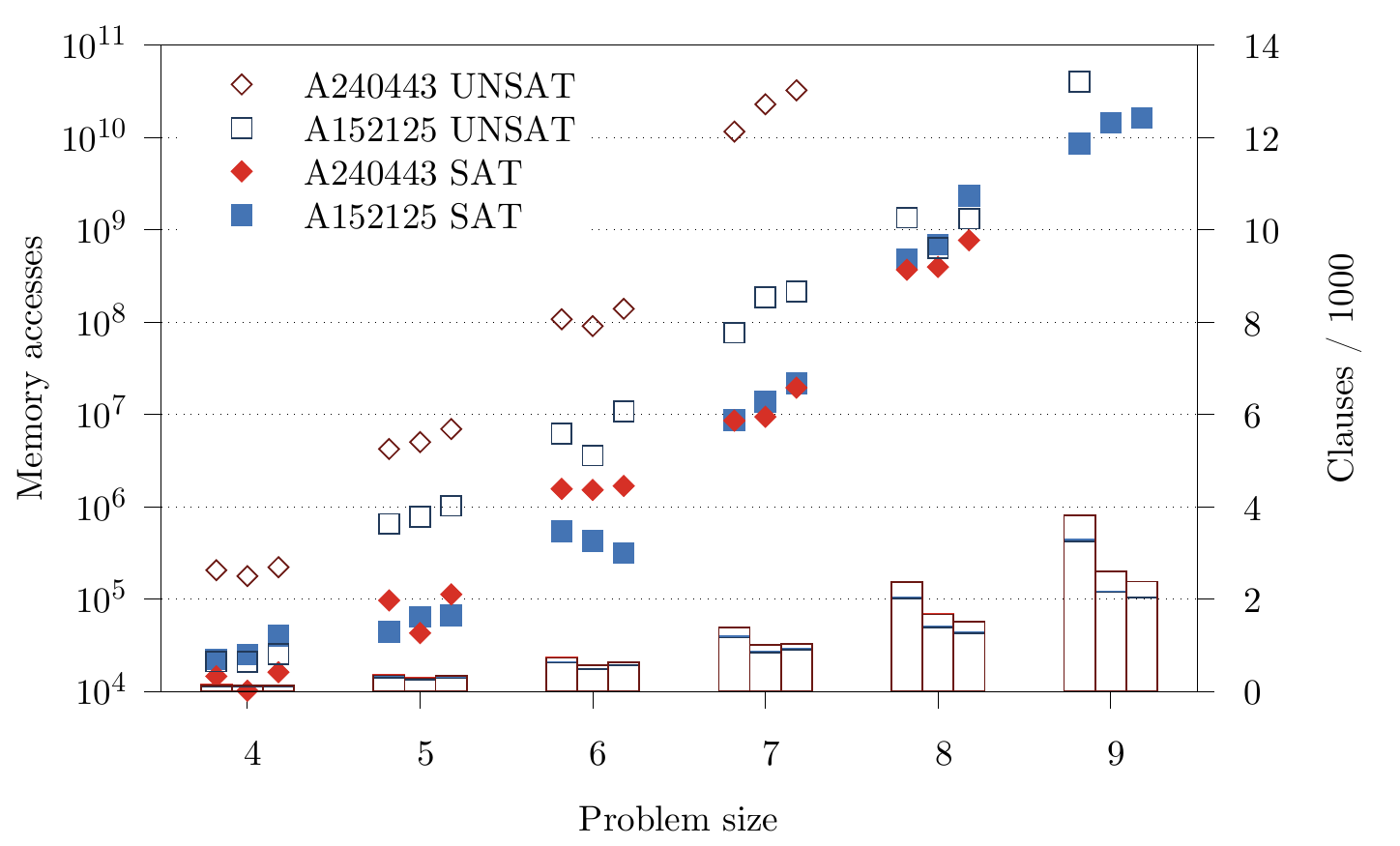}}}
{\scalebox{1.0}{\includegraphics{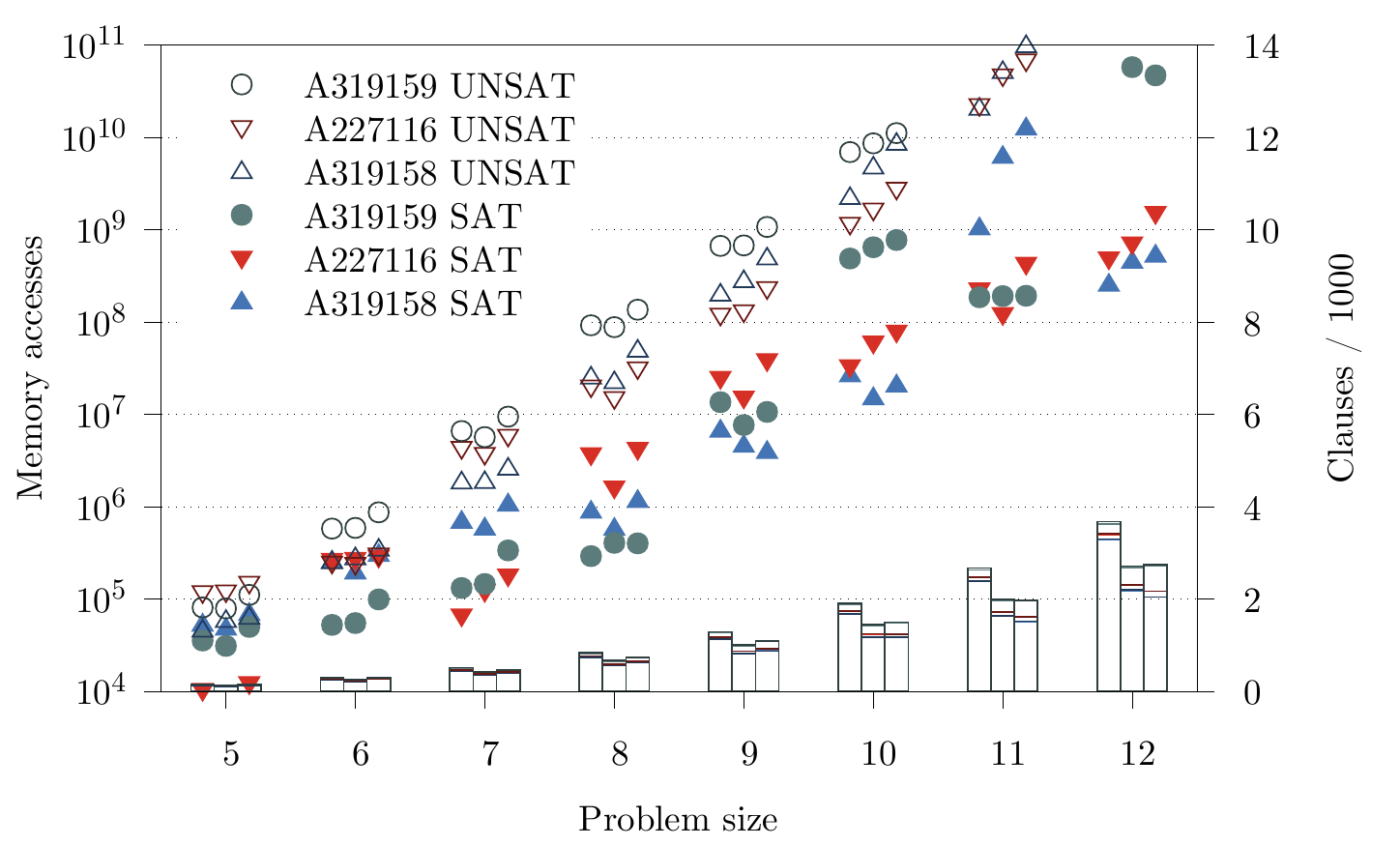}}}
	\caption{Comparison between results for different encodings.  Top: test cases involving squares; bottom: test cases involving triangles.  The numbers of memory accesses are plotted as symbols, using the logarithmic scale on the left.  The numbers of clauses are plotted as bars, using the linear scale on the right. At each size, the encodings are, from left to right: sequential-counter, tree-based and sort-based.  For each encoding approach, the unstrengthened variant is used.}
	\label{fig:compare}
	\end{figure}

In Sections~\ref{sec:sinz}, \ref{sec:bail} and~\ref{sec:abio}, several variants of each encoding are presented.  The variants for each are compared in Figures~\ref{fig:sinzvariants}, \ref{fig:bailvariants} and~\ref{fig:abiovariants}.  The encodings are compared with each other in Figure~\ref{fig:compare}.  The following conclusions can be drawn from these results:
\begin{itemize}
  \item The UNSAT cases require considerably more effort than the corresponding SAT cases -- often by 2 or 3 orders of magnitude.
  \item The growth rate of mem count with problem size is rapid.  There are occasional anomalies -- for example, in A319158~SAT, all encodings solved $L=12$ faster than $L=11$.
  \item When the sequences detailed in Section~\ref{sec:testcases} are used to define test cases for the encodings and solvers considered here, using modest computing power, suitable problem sizes are smaller than the largest known values in the OEIS.  (To determine a value in the sequence, adjacent SAT and UNSAT values must be solved.)  Because of the rapid growth with problem size, the current methods do not appear to be suitable for extending the known values in these sequences.  Problems consisting of less than 10000 clauses can be challenging.  In fact, the test cases here, though motivated by combinatorial searches, show some resemblances with sets of clauses that are deliberately crafted to be difficult to solve~\cite{Spence:2015}.
  \item The `broader' problems, A240443$(L)$ and A319159$(L)$, which consider more squares and triangles (respectively) than their counterparts, use only slightly more clauses but require substantially more effort to solve, particularly for the UNSAT cases.  However, this trend is reversed when comparing A227116$(L)$ and A319158$(L)$: A227116$(L)$ is broader, and has slightly more clauses, but often requires less effort.
  \item The effect of strengthening each encoding is small compared to the other effects.  The fully-strengthened sort-based encoding runs slightly faster than the basic variant: if we exclude test cases whose median mem count was less than $10^8$, the fully-strengthened variant was faster in 20 cases out 23, but the differences are generally small.  In fact, the lack of effect is much more surprising, because strengthening requires a substantial increase in the number of clauses for each encoding.
  \item For the majority of the test cases shown, the sequential-counter encoding is faster than the tree-based encoding.  This conclusion becomes much stronger when the smallest test cases are eliminated: for problems that required a median mem count greater than $10^8$, the sequential-counter encoding had the lowest median mem count in 18 cases out of 21.  This conclusion is not sensitive to the arbitrary limit of $10^8$.  The same conclusion is drawn if the fastest repeat from each set is used instead of the median.
  \item The sort-based encoding appears to be slower than the other encodings for almost all test cases here.  The pairwise variant of sort-based encoding~\cite{Codish:2010} has been  tested; for the current test cases, this appears to be slower than the mergesort variant~\cite{Abio:2013} discussed in Section~\ref{sec:abio}.  This contradicts results on other test cases~\cite{Codish:2010}, so further investigation may be beneficial.
  \item These comparisons accentuate the well-known observation that the encoding size is not a reliable indicator of problem difficulty, even in very closely related problems.  The points above give examples where problem effort increases by orders of magnitude for a small change in clause count, and other examples where the problem effort is almost unchanged even when clause count is doubled.  The same conclusion applies to the total numbers of variables and other measures of the size of a \Sat problem.  The insensitivity of runtime to encoding size may have implications for optimization or preprocessing schemes such as Bounded Variable Addition~\cite{Manthey:2013}, where encoding size (such as total variables and clauses) may be used as a proxy for problem difficulty.  Many publications assume that a smaller encoding is better, perhaps implying faster runtimes, but this is not inevitable.  In fact, in Figure~\ref{fig:compare}, solution effort is anti-correlated with encoding size when encoding methods are compared at a fixed size.  The fastest method, the sequential counter, has the largest numbers of clauses, auxiliary variables and literals; this is exemplified in Table~\ref{tab:counts}.
  \item It is interesting, and perhaps disconcerting, that a contest between the encodings would give different conclusions for smaller problems and for larger problems in the range considered here.  This of course raises the question of whether the conclusions would change again for problem sizes beyond this range.  Similarly, the current work has only added to the contradictory conclusions of previous work mentioned in Section~\ref{sec:intro}. The overall conclusion must be that testing is required to select the best encoding method for a specific problem.
\end{itemize}

\begin{figure}[htp] 
	\centering
{\scalebox{1.0}{\includegraphics{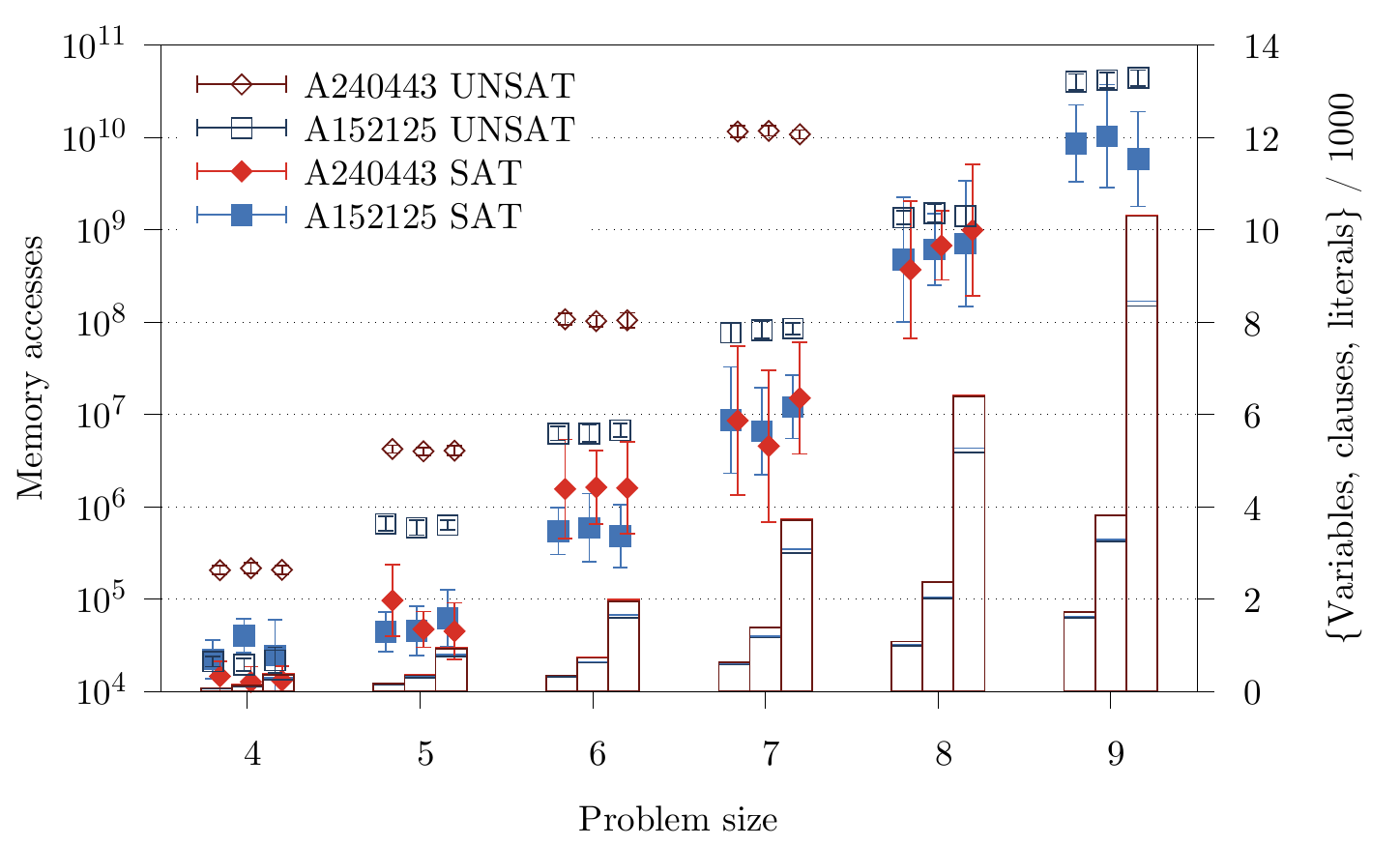}}}
{\scalebox{1.0}{\includegraphics{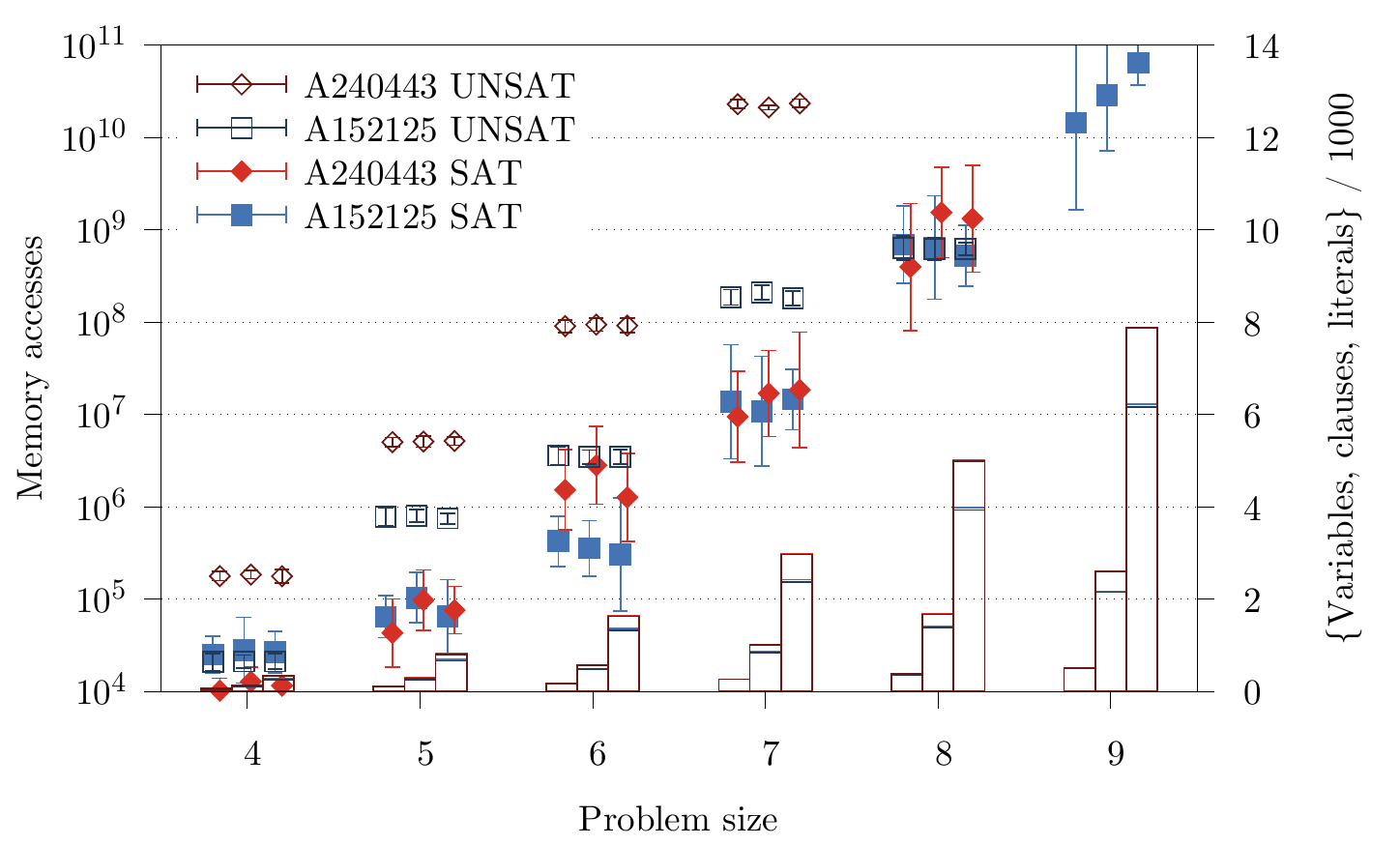}}}
	\caption{Comparison between results for different encodings when the main variables are reordered in the cardinality constraint.  There is no significant impact on the mem counts. Top: unstrengthened sequential-counter encoding; bottom: unstrengthened tree-based encoding.  Symbols and axes are as in Figure~\ref{fig:compare}~(top), except that the bars at each size represent (from left to right) the numbers of variables, clauses and literals in the encodings -- these counts are identical for the three orderings. The error bars are defined in Section~\ref{sec:methodology}.  At each size, the orderings of main variables are, from left to right: each row of the square grid, in order; a square spiral starting at the centre of the grid; a single random permuation of all points.}
	\label{fig:orderings}
	\end{figure}

To explain the differences between the runtimes of the encoding methods, one hypothesis is that it could be significant which main variables are closely connected to each other.  The variables are presented in the same order to the cardinality constraints, which then interconnect them in different ways.  Perhaps the interconnections explain the differences?  This hypothesis is rejected by Figure~\ref{fig:orderings}, where the variables are encoded into the cardinality constraints in various orders.  The results from each cardinality method appear to be unchanged.  (Not all test cases and methods are shown in the figure, but the conclusion applies throughout.)  These results are a suitable opportunity to display the error bars on the results; as discussed in Section~\ref{sec:methodology}, these are representative of the dispersion of single values whose medians are plotted in Figures~\ref{fig:sinzvariants} to~\ref{fig:abiovariants}.

In cases using the sequential-counter constraint, Marques-Silva and Lynce \cite{MarquesSilva:2007} found substantial improvements when the solver was adapted to choose only main variables for branching decisions.  The solver in that study was MiniSat.  From the perspective of strengthening, this finding makes some sense: unstrengthened encodings of a constraint allow the auxiliary variables to adopt `unhelpful' solutions, which may temporarily enforce stronger constraints than required.  If only main variables are used for decisions, these unhelpful solutions are not explored.  However, a similar adaptation to the {\sc sat13} solver was tried in the current study, but here it increased solver effort rather than decreased it.

This paper has developed strengthened variants of encodings for cardinality constraints.  These variants are recommended when the task is to enumerate all the main-variable solutions to a problem.  In the test cases considered, however, they have surprisingly small effects on the solver effort required to reach a single solution.

\end{document}